\documentclass[prl,twocolumn,english,superscriptaddress,floatfix,aps,10pt,balancelastpage,footinbib]{revtex4-2}

\usepackage{stylesheet}
 
\begin{document}


\title{Higher moment theory and learnability of bosonic states}
\begin{CJK}{UTF8}{gbsn}

\author{Joseph~T.~Iosue}
\email{jtiosue@gmail.com}
\affiliation{\QUICS}
\affiliation{\JQI}

\author{Yu-Xin Wang (王语馨)}
\affiliation{\QUICS}

\author{Ishaun Datta}
\affiliation{Department of Computer Science, Stanford University, Stanford, California 94305, USA}

\author{Soumik Ghosh}
\affiliation{Department of Computer Science, University of Chicago, Chicago, Illinois 60637, USA}

\author{Changhun Oh}
\affiliation{Department of Physics, Korea Advanced Institute of Science and Technology, Daejeon 34141, Korea}

\author{Bill Fefferman}
\affiliation{Department of Computer Science, University of Chicago, Chicago, Illinois 60637, USA}

\author{Alexey~V.~Gorshkov}
\affiliation{\QUICS}
\affiliation{\JQI}

\date{\today}

\begin{abstract}
We present a sample- and time-efficient algorithm to learn any bosonic Fock state acted upon by an arbitrary Gaussian unitary. As a special case, this algorithm efficiently learns states produced in Fock state BosonSampling, thus resolving an open question put forth by Aaronson and Grewal \cite{aaronson2023efficient-tomog}. 
We further study a hierarchy of classes of states beyond Gaussian states that are specified by a finite number of their higher moments.
Using the higher moments, we find a full spectrum of invariants under Gaussian unitaries, thereby providing necessary conditions for two states to be related by an arbitrary (including active, e.g.~beyond linear optics) Gaussian unitary.
\end{abstract}

\maketitle
\end{CJK}


%
The intrinsic exponential sample and time complexity of quantum state tomography as system size scales \cite{bruss1999optimal, o2016efficient, haah2016sample, kueng2017low, chen2023does} motivates the search for natural classes of quantum states that can be learned efficiently with a small number of samples and modest computation. Existing learning algorithms often rely crucially on classical simulability (e.g.~\cite{cramer2010efficient, montanaro2017learning}). Do there exist efficient learners even when classical simulation is hard, such as for quantum advantage experiments? While prior results address this question for some states generated by IQP circuits \cite{arunachalam2022optimal} and quantum circuits at sufficiently low depth \cite{fefferman2024anti, huang2024learning, landau2025learning}, this question has remained open for Fock states acted upon by Gaussian unitaries, as in BosonSampling \cite{aaronson2011the-computation}. 
In standard BosonSampling experiments \cite{doi:10.1126/science.1231692,doi:10.1126/science.1231440,Tillmann2013-gg,Crespi2013-ei}, an $n$-mode Fock state
(each mode containing zero or one photon)
is acted upon by a linear optical unitary.
Throughout this work, we consider more general states---namely arbitrary Fock states acted upon by an arbitrary Gaussian unitary.
We ask:
\textit{Does there exist a sample- and time-efficient algorithm to learn any such state?}
In this work, we answer this question in the affirmative, thereby resolving an open problem posed by Aaronson and Grewal \cite{aaronson2023efficient-tomog}.

We study higher moments (that is, beyond second moments) and their applicability to finding Gaussian invariants and to learning quantum states.
Given a bosonic state, using its moments, we find all polynomials in the moments that are invariant under the action of a Gaussian unitary.
We define classes of states that are fully specified by their first $t$ moments.
We show that Fock states acted upon by an arbitrary Gaussian unitary are fully defined by their first four moments.
Using this, we derive an explicit sample- and time-efficient algorithm for learning such states.
We then compare to other learning algorithms \cite{aaronson2023efficient-tomog,bittel2025energy-independ,mele2024learning-quantu,zhao2024complexity-of-q}, show natural extensions of our algorithm, discuss implications for finding states that are intrinsically \textit{hard} to learn, and suggest future directions.
Our algorithms establish a new framework for learning quantum states that could have implications for other classes of bosonic states as well as other non-bosonic systems.

\textit{Theory of higher moments.---}%
Second moments in bosonic and fermionic quantum information theory have been an incredibly fruitful area of study, and indeed Gaussian states are fully specified by just their first and second moments \cite{weedbrook2012gaussian-quantu,serafini2017quantum-continu,hackl2021bosonic-and-fer}.
Beyond Gaussian states, \emph{any} bosonic state is completely characterized by its moments \cite[Sec.~3.8.1]{welsch1999ii-homodyne-det}.
As such, higher moments and cumulants have been studied
\cite{xiang2018evaluation-of-t,cardin2024photon-number-m,jiang2010the-cumulants-o}.
In this work, we will deal extensively with the \emph{moment tensors} of a state \( \rho \),
\begin{equation}
    \Sigma^{(t)}_{i_1,\dots,i_{t}} =
    \Tr\bargs{\rho \tilde r_{i_1} \dots \tilde r_{i_{t}} }
    ,
\end{equation}
where we defined the quadrature operators \( \bm r = (x_1,\dots,x_n,p_1,\dots,p_n) \) and the centralized operators \( \tilde r_i = r_i - \Tr[\rho r_i] \).
\( \Sigma^{(t)} \) contains the information about all \( t \)-degree moments.
When $t = 2s$ for an integer $s$, we consider a reshaping of $\Sigma^{(t)}$, which we call $\Lambda^{(s)}$, that is an operator on \( (\bbC^{2n})^{\otimes s} \), and thus
a \( (2n)^{ s} \times (2n)^{s} \) matrix~\footnote{Throughout this work, we assume that in any such reshaping, the first $n$ indices in the  $\Sigma^{(t)}$ tensor act as the row indices of $\Lambda^{(s)}$, and the rest of the indices correspond to the column induces.}.


A mixed bosonic Gaussian state on \( n \) modes can be defined uniquely as the maximal (von-Neumann) entropy state with a given covariance matrix
\cite{surace2022fermionic-gauss,jaynes1957information-the,jaynes1957information-the-II}.
A straightforward generalization of the calculation for classical probability distributions \cite[Thm.~12.1.1]{cover2005elements-of-inf} yields the more standard definition that a Gaussian mixed state is a thermal state of a Hamiltonian that is quadratic in the quadrature operators
\cite{serafini2017quantum-continu}.
A pure Gaussian state is then defined as the limit of a mixed Gaussian state as the temperature is taken to \( 0 \).
Given a set of operators $F_i$, the maximal entropy state subject to constraints on the expectation value of each $F_i$ is proportional to the exponential of a linear combination of the $F_i$
\cite{jaynes1957information-the-II}.
Motivated by the fact that
any bosonic state
\footnote{We will always only consider physical bosonic states that have finite moments of every order. Technically, we restrict our attention to pure states that live in Schwartz space \( S(\bbR^n) \)and mixed states that are analytic Schwartz operators \cite[Def.~4.1]{keyl2016schwartz-operat}.}
is completely characterized by its moments \cite[Thm.~4.4]{keyl2016schwartz-operat}\cite[Sec.~3.8.1]{welsch1999ii-homodyne-det}, we define a bosonic \Gt mixed state to be the maximal entropy state subject to constraints on its first \( t \) moments.
An application of Ref.~\cite{jaynes1957information-the-II} in the mixed state case (see the End Matter for the pure state case) then yields the following equivalent definition.

\begin{definition}
    A bosonic mixed state on \( n \) modes is called a \emph{\Gt state} if it is the thermal state of a degree \( \leq t \) Hamiltonian in the quadrature operators \( \bm r \).
    A pure state is a \Gt state if it is the ground state of a non-degenerate degree \( \leq t \) Hamiltonian.
    The \emph{Gaussian degree} of a bosonic state is the minimum \( t \) such that it is a \Gt state.
\end{definition}

To avoid peculiarities stemming from unboundedness (cf.~Ref.~\cite{cover2005elements-of-inf}), we consider only \Gt states for even \( t \).
A \Gt state is fully specified by its first $t$ moments in the following sense (see the End Matter).
Suppose Alice gives Bob $\Sigma^{(t')}$ for all $t' \leq t$, and Alice promises Bob that those moments came from a \Gt state.
Then Bob has enough information to completely reconstruct the state.
Thus, Bob can, for example, compute $\Sigma^{(t+s)}$ for any $s$.
In the case of $t=2$, this is precisely what happens for Gaussian states---if Alice gives Bob a mean vector and a covariance matrix and promises that they came from a Gaussian state, then Bob can completely reconstruct the state.

%
%

The first obvious feature of the set of \Gt states is that it is closed under Gaussian unitaries. 
This is an immediate consequence of the linearity of Gaussian transformations---namely that
a Gaussian unitary \( \calU_S \) is specified by a symplectic matrix \( S \in \Sp(2n, \bbR) \) and displacements $d_i\in\bbR$, and acts as
\( \calU_S^\dag \tilde r_i \calU_S = \sum_j S_{ij} \tilde r_j \) \cite{serafini2017quantum-continu}.
Thus, the Gaussian unitary \( \calU_S \) acts on the moments as
\begin{equation}
    \label{eq:symplectic-action}
    \Sigma^{(t)} \mapsto S^{\otimes t} \Sigma^{(t)},
    \qquad 
    \Lambda^{(t)} \mapsto S^{\otimes t} \Lambda^{(t)} (S^T)^{\otimes t}
    .
\end{equation}
In other words, different degree moments do not mix under Gaussian unitaries.
It follows that the Gaussian degree of a bosonic state is invariant under Gaussian unitaries, and, analogously to \cite[Thm.~3]{chabaud2020stellar-represe} regarding the stellar rank of a state,
a unitary is Gaussian if and only if it always leaves the Gaussian degree invariant.

\textit{A spectrum of symplectic invariants and Gaussian convertibility.---}%
In the realm of Gaussian states,
the correspondence between Gaussian unitaries and symplectic matrices has been very useful.
In particular, it allows for any state to be specified by a unique \emph{normal form} via Williamson's diagonalization \cite{serafini2017quantum-continu}.
Specifically, we can generate any Gaussian state by acting a certain Gaussian unitary on a product state that can be written as the thermal state of a quadratic Hamiltonian purely consisting of linear combinations of single-mode number operators.
The product thermal state is uniquely specified (up to permutations of the modes) by the \emph{symplectic eigenvalues} of the covariance matrix.
The covariance matrix is defined by \( V = \Re\Lambda^{(1)} \), and the symplectic eigenvalues are the positive eigenvalues of \( \i \Omega V \), where \( \Omega \) is the corresponding symplectic form encoding the canonical commutation relations between quadrature operators \cite{serafini2017quantum-continu}.
The symplectic eigenvalues of the covariance matrix can be defined for any bosonic states, although the correspondence to a product thermal Gaussian state only applies if the original state is Gaussian. Those eigenvalues are examples of \emph{symplectic invariants}, meaning that they are unchanged under the application of a Gaussian unitary.
More precisely, in order for a quantity to be invariant under Gaussian unitaries, it must be a symplectic invariant \textit{and} be invariant under displacements. Because we defined $\Sigma^{(t)}$ to be central moments, it is automatically invariant under displacements.

%
%
For a permutation on \( t \) elements \( \pi \in S_t \), define the operator \( W_\pi \) on \( (\bbC^{2n})^{\otimes t} \) as \( W_\pi \ket{i_1,\dots,i_t} = \ket{i_{\pi(1)}, \dots, i_{\pi(t)}} \). 
Define the vector $\theta \in (\bbC^{2n})^{\otimes 2}$ by $\theta_{ij} = \Omega_{i,j}$.
Finally, for any tuple of positive integers $\bm s = (s_1,\dots,s_k)$, define $\abs{\bm s} = \sum_i s_i$, and
let $\Gamma^{(\bm s)} \in  (\bbC^{2n})^{\otimes \abs{\bm s}}$ be $\bigotimes_i \Sigma^{(s_i)}$.
Define $\lambda(\rho)$ to be the collection of symplectic invariants $\bra{\theta^{\otimes \abs{\bm s}/2}} W_\pi \ket{\Gamma^{(\bm s)}}$
for all $\bm s$ whenever $\abs{\bm s}$ is even (note there are infinitely many) and all $\pi \in S_{\abs{\bm s}}$.

Suppose that \( Q(\rho) \) is a function that is invariant under all Gaussian unitaries, meaning that \( Q(U\rho U^\dag) = Q(\rho) \) for all Gaussian unitaries \( U \).
Further, suppose that $Q$ is a polynomial in the entries of the moment tensors; such a polynomial is natural to consider given that a state is fully specified by its moments.
Because Gaussian unitaries act via a tensor product representation on the moment tensors (cf.~\cref{eq:symplectic-action}), 
it follows from 
Ref.~\cite[Sec.~5]{goodman2009symmetry-repres}
that \( Q(\rho) \) can be written as a function of exclusively the symplectic invariants \( \lambda(\rho) \).

While all polynomial invariants can be expressed in terms of $\lambda(\rho)$, we highlight another particularly nice set of invariants that we denote $\lambda_E(\rho)$.
Notice that $\Gamma^{(\bm s)}$ can be reshaped into a square matrix $\bar\Gamma^{(\bm s)}$ of dimension $(2n)^{\abs{\bm s}/2}$.
For a permutation $\pi \in S_{\abs{\bm s}/2}$, define $\lambda_E(\rho)$ to be the collection of eigenvalues of $(\i \Omega)^{\otimes \abs{\bm s} / 2} W_\pi \bar\Gamma^{(\bm s)}$.
By examining the characteristic polynomial of the matrix and using the defining property $S\Omega S^T = \Omega$ of a symplectic matrix, one can indeed verify that the elements of $\lambda_E(\rho)$ are symplectic invariants.
Every element of $\lambda_E(\rho)$ can be expressed as a polynomial in the elements of $\lambda(\rho)$ via Girard–Newton formulae, but the reverse is not obviously true.
In other words, $\lambda(\rho)$ generates the full set of polynomial invariants, while $\lambda_E(\rho)$ generates a subset.
Nevertheless, $\lambda_E(\rho)$ is a natural subset to consider seeing as it consists of natural generalizations of the symplectic eigenvalues of the covariance matrix.

One application of the symplectic invariants is the Gaussian convertibility problem.
Following e.g.~Refs.~\cite{chabaud2020stellar-represe,hahn2024assessing-non-g}, two states are said to be Gaussian convertible if there exists a Gaussian unitary taking one to the other.
Whether two states are Gaussian convertible is determined by the equivalence or inequivalence of all of their respective symplectic invariants.
Thus, in order to establish that two states are not Gaussian convertible, it suffices to find a single element (i.e.~witness) in $\lambda(\rho)$ (any element of $\lambda_E(\rho)$ of course also suffices)
that is different.
Previous methods of finding invariants for $n$-mode bosonic states have primarily focused on only passive Gaussian unitaries acting within the Hilbert space of a finite Fock number cutoff \cite{parellada2023no-go-theorems-,migdal2014multiphoton-sta,draux2025invariants-in-l}, and moreover many of the other invariants \cite{draux2025invariants-in-l,upadhyaya2025majorization-th} are often computationally demanding for large numbers of modes.
In contrast, the invariants in $\lambda(\rho)$ require no Fock space cutoff and allow for general Gaussian unitaries.
Furthermore,
one can compute $\poly{n}$ of these invariants in $\poly{n}$ time by enumerating all the invariants coming from $\bm s$ with a constant cutoff on $\abs{\bm s}$.



It is not immediately obvious whether two states having the same $\lambda(\rho)$ is a sufficient condition to be related by a Gaussian unitary.
We discuss this more in the End Matter.


On a single mode, the Fock state \( \ket1 \) and the photon subtracted squeezed state \( \propto a\ket\xi \) are Gaussian convertible \cite[Supp.~VB]{chabaud2020stellar-represe} \cite{PhysRevA.55.3184,PhysRevLett.92.153601,Pasharavesh:24,10.1002/qute.202400616}.
Note that \( \ket1 \) is a \G*4 state because it is the ground state of the Hamiltonian \( H = (a^\dag a-1)^2 \), and indeed all the symplectic invariants agree with the photon subtracted squeezed state.

Outside of fixed boson number subspaces, the two-mode states \( \propto \ket{22} + \sqrt{3}\ket{10} + \sqrt{2} \ket{01} \) and \( \propto \ket{22} + \sqrt{1}\ket{10} + \sqrt{4} \ket{01} \)  have different symplectic invariants coming from the eigenvalues of \( (\i\Omega)^{\otimes 2}W_\pi\Lambda^{(2)} \), thus proving that they cannot be related by a Gaussian unitary.
Note that the symplectic eigenvalues of the covariance matrices---i.e.~the eigenvalues of \( \i\Omega \Lambda^{(1)} \)---are the same for both these states, thus providing an example of the necessity of considering higher moments in the Gaussian conversion problem.


\textit{Learning states.---}%
Motivated by the fact that a finite set of moments fully describe a \Gt state, we now derive an explicit learning (a.k.a.~tomography) algorithm for a class of \G*4 states.
As described in the previous section, Fock states are \G*4 states and thus are fully specified by their second and fourth moment matrices (odd moments vanish).
We consider Fock states acted on by arbitrary Gaussian unitaries.
Such states are of great recent interest due to their relevance in BosonSampling~\cite{aaronson2011the-computation}.
Given sample access to a state \( \calU_S \ket{\bm f} \) for an unknown Fock state \( \bm f = (f_1,\dots f_n) \) on \( n \) modes and an unknown Gaussian unitary $\calU_S$ specified by the symplectic matrix \( S \in \Sp(2n, \bbR) \) (we assume zero displacements, as they can easily be learned by measuring first moments),
we can make measurements in order to build approximations to \( \Lambda^{(1)} \) and \( \Lambda^{(2)} \), where \( \Lambda^{(t)} \) is a \( (2n)^t \times (2n)^t \) matrix.
In particular,
the matrix elements of \( \Lambda^{(t)} \) can be estimated to a given error using Gaussian (e.g.~homodyne or heterodyne) measurements \cite[Sec.~3.8.1]{welsch1999ii-homodyne-det} \cite{richter1996determination-o,wunsche1997radon-transform}.
For constant $t$, achieving inverse polynomial precision in the estimate of $\Lambda^{(t)}$ can be done with polynomially many measurements
(see Propositions S14 and S17 of the Supplemental Material).
Once the moments are known to a given precision, we develop an efficient algorithm to learn the state \footnote{Our goal is to learn an efficient representation of $\vert\psi'\rangle$ such that $\vert\psi'\rangle$ has high overlap with the true state $\vert\psi\rangle$, so that $\langle\psi\vert\psi'\rangle \geq 1 - \varepsilon$. We are interested in the succinct representation $\calU_{Q}\vert\bm g\rangle$ of $\vert\psi'\rangle$.}.

\begin{theorem}
    \label{thm:main-thm-gaussian}
    Let \( \ket\psi = \calU_S \ket{\bm f} \) for an unknown symplectic matrix \( S \in \Sp(2n, \bbR) \) specifying an arbitrary Gaussian unitary (modulo displacements) and an arbitrary Fock state \( \ket{\bm f} \).
    If our measurements \( \Lambda^{(1)\prime},\Lambda^{(2)\prime} \) of the moment matrices \( \Lambda^{(1)},\Lambda^{(2)} \) satisfy \( \norm*{\Lambda^{(t)\prime} - \Lambda^{(t)}} \leq \varepsilon_t \),
    then we can efficiently find a \( Q \in \Sp(2n, \bbR) \) and \( \bm g \) such that
    \begin{equation}
        \abs*{\bra{\bm f}\calU_S^\dag \calU_Q \ket{\bm g}} 
        \geq 
        1
        - \bigO{
        \varepsilon_1 n^2 \e^{3s}  f_{\rm max}^3 + \varepsilon_2 n^2 \e^{6s} f_{\rm max}
        }
        ,
    \end{equation}
    where \( f_{\rm max} = \max_i f_i \) and \( s \) is the maximum magnitude of squeezing in \( S \) (that is, \( \e^s \) is the largest singular value of \( S \)).
\end{theorem}

The special case of \cref{thm:main-thm-gaussian} when restricting to passive Gaussian unitaries (\textit{a.k.a.}~linear optics, where $s = 0$) resolves an open question put forth by Aaronson and Grewal \cite{aaronson2023efficient-tomog}.
Given this restriction, in the Supplemental Material, we track the constant factors to arrive at explicit, non-asymptotic bounds.
One can in principle also track the constant factors in the proof of \cref{thm:main-thm-gaussian} for arbitrary Gaussian unitaries, but we do not do it in this work.
\tempignore{
Finally, 
in Corollaries S15, S16, and S18 of the Supplemental Material, we consider the full end-to-end learning algorithm and derive bounds on the number of measurements from the state that are needed in order to learn the state to a desired fidelity.
}
In order to understand the full end-to-end learning algorithm,
the only remaining ingredient beyond \cref{thm:main-thm-gaussian} is to determine how many measurements are needed in order to estimate the moment matrices to the desired precision.
This depends on the measurement technique.
In this work, we will examine heterodyne measurements, but emphasize that \cref{thm:main-thm-gaussian} can be applied to the moment matrices however they are computed.

\ignore{
\note{rewrite this}
In particular, given $N$ copies of $\ket\psi$, the state is sampled via e.g.~homodyne measurements, yielding estimates $\Lambda^{(t)\prime}$ of the moment matrices $\Lambda^{(t)}$ for $t=1,2$ \cite[Sec.~3.8.1]{welsch1999ii-homodyne-det} \cite{richter1996determination-o,wunsche1997radon-transform}.
Running the algorithm in \cref{thm:main-thm-gaussian} with $\Lambda^{(t)\prime}$, we learn $\ket\psi$ to fidelity $1-\gamma$ with probability $1-\delta$, where $\delta > 0$ because of the probabilistic nature of measuring $\Lambda^{(t)\prime}$.
We want to know: if we desire $\delta = \bigO{1/n^\beta}$ and $\gamma = \bigO{1/n^{\alpha}}$ for fixed constants $\alpha,\beta > 0$, what is the required $N$?
In Corollaries S14, S15, and S17, by applying \cref{thm:main-thm-gaussian}, we show that the required $N$ can be upper bounded as a polynomial in $n$, $f_{\rm max}$, and $\e^s$, and we also show explicit numerical simulations showing the expected number of heterodyne samples needed to achieve a desired fidelity.
}

In the Supplemental Material, we show that given $N$ copies of $\ket\psi$, using heterodyne measurements, we can produce estimates $\Lambda^{(t)\prime}$ of the moment matrices $\Lambda^{(t)}$ for $t=1,2$ to $1/\poly N$ precision with high probability.
Applying this to \cref{thm:main-thm-gaussian}, for any desired success probability $1-\delta$, $N \sim \poly{n, \e^s, f_{\rm max}, 1/\delta, 1/\gamma}$ suffices to learn the state to fidelity $1-\gamma$.

\begin{corollary}
    \label{cor:main-result}
    Let $S \in \Sp(2n, \bbR)$ specify an unknown arbitrary Gaussian unitary $\calU_S$ and $\ket{\bm f}$ be a Fock state. 
    Then
    $N \sim \bigO{n^8 f_{\rm max}^{10} \e^{20s} 
    \log(n/\delta) / \gamma^2}$
    heterodyne samples from the state $\calU_S\ket{\bm f}$ suffices (via our efficient classical algorithm) to learn a $Q\in \Sp(2n,\bbR)$ and $\bm g$ satisfying $\abs*{\bra{\bm f}\calU_S^\dag \calU_Q\ket{\bm g}} \geq 1 - \gamma$ with probability at least $1-\delta$.
\end{corollary}

\ignore{
In practice, how does one apply this? It depends on the setting.
In one setting, we could be promised that $f_{\rm max}$ and $\e^s$ are bounded by a known constant, thus giving us a way to determine $N$ solely in terms of $n$.
In another setting, suppose that we are not promised that $f_{\rm max}$ and $\e^s$ are bounded.
In this case, in order to choose $N$, we may first need to measure from $\ket\psi$ to upper bound $f_{\rm max}$ and $\e^s$.
As an example, we first consider the case when $s=0$ (i.e.~the BosonSampling setting of a Fock state acted upon by a linear optical unitary).
In this case, the state $\ket\psi$ we are trying to learn is an eigenstate of the total boson number operator. Thus, we can perform a standard BosonSampling measurement to detect all bosons and learn that the state has $B$ bosons. Because $f_{\rm max} \leq B$, we see that the required $N$ is upper bounded by a polynomial in $n$ and $B$, and we can choose a sufficient $N$ given that we know $n$ and $B$.
In fact, in typical Boson sampling, $f_{\rm max} = 1$ \cite{aaronson2011the-computation}, so that measuring $B$ is not even necessary.
\tempignore{, and Corollary S14 automatically tells us the required $N$ in terms of only $n$.}
}

The specific problem put forth by \cite{aaronson2023efficient-tomog} was for $f_{\rm max} =1$ and $s = 0$ -- for this problem, our algorithm therefore requires $\sim n^8\log n$ samples.
We give numerical evidence in the Supplemental Material that, at least for random unitaries, the scaling is closer to $n^6$ in practice.
Finally, in the case when it is not a priori known that $\e^s$ is bounded by a constant, we expect that $\e^s$ and $f_{\rm max}$ can be upper bounded using the knowledge of the covariance matrix $\Re\Lambda^{(1)}$ of $\ket\psi$. Specifically, $\e^s$ and $f_{\rm max}$ control the energy density of the state. The largest symplectic eigenvalue of $\Re\Lambda^{(1)}$ and the largest squeezing strength of the symplectic transformation that diagonalizes $\Re\Lambda^{(1)}$ correspond to $\e^s$ and $f_{\rm max}$, respectively. In an end-to-end algorithm, we can thus rigorously bound $\e^s$ and $f_{\rm max}$ through first measuring the covariance matrix. This can be done with polynomial number of samples~\cite{zhao2024complexity-of-q}. Our algorithm still requires the promise that the covariance matrix has finite matrix elements, an assumption that is satisfied in typical physical experiments.
We further suspect that the adaptive measurement technique from Ref.~\cite{bittel2025energy-independ} can be applied to change the $\poly{\e^s}$ dependence to a $\poly{s}$ dependence.

We emphasize that, given the moment matrices, the algorithm in \cref{thm:main-thm-gaussian} runs efficiently and only uses basic linear algebra routines---namely, matrix diagonalization, singular value decomposition (SVD), and Williamson decomposition.
Motivated by numerical simulations in the Supplemental Material, we
suspect that the bounds in \cref{thm:main-thm-gaussian} are loose, and that in practice the degrees of the polynomial dependencies on \( \e^s, f_{\rm max} \), and \( n \) are smaller than stated in the theorem.
%
While the full algorithm, theorem statement, and proof are provided in the Supplemental Material,
here we provide a high-level overview of a special case of the algorithm that nonetheless provides good intuition.
We will assume that the moment matrices are known perfectly (i.e.~\( \varepsilon_1=\varepsilon_2=0 \)).
We will further assume that the initial Fock state \( (f_1,\dots,f_n) \) is \( b,\dots,b \) for a fixed integer \( b\geq 1 \).
Finally, we will restrict our focus to passive Gaussian unitaries, which are specified by an element \( S \in K(n) = \Sp(2n,\bbR) \cap \O(2n) \) \cite{serafini2017quantum-continu}.
Because \( K(n) \) is isomorphic to the unitary group \( U(n) \), we will denote the corresponding \( n\times n \) unitary by \( W \) and the Gaussian unitary as \( \calU_W \).
Thus, we wish to learn an unknown unitary \( W \) from the moment matrices of the state \( \ket\psi = \calU_W \ket{b\dots b} \).
It follows that we can restrict our attention to the moments
\begin{equation}
    \sigma^{(t)}_{i_1,\dots,i_t; j_1,\dots,j_t} = \bra\psi a_{i_1}\dots a_{i_t}a^\dag_{j_1}\dots a^\dag_{j_t}\ket\psi ,
\end{equation}
where \( \calU_W \) acts on the annihilation operators \( a_1,\dots,a_n \) as \( \calU_W^\dag a_i \calU_W = \sum_{j=1}^n W_{ij}a_j  \) \cite{serafini2017quantum-continu}.
We view \( \sigma^{(t)} \) as an \( n^t \times n^t \) matrix; that is, \( \sigma^{(t)} \) is an operator on \( (\bbC^n)^{\otimes t} \).
By construction, \( \calU_W \) acts as \( \sigma^{(t)} \mapsto W^{\otimes t} \sigma^{(t)}W^{\dag\otimes t} \).

We first consider the fourth moment matrix for the state \( \ket{b\dots b} \), and we denote it by \( \sigma^{(2)}_0 \). Because \( \sigma^{(2)}_0 \) is a matrix on \( (\bbC^{n})^{\otimes 2} \), we can represent it in bra-ket notation using the standard basis \( \ket{1},\dots,\ket{n} \) of \( \bbC^n \).
Some algebra shows that
\begin{equation}
    \sigma^{(2)}_0 = (b+1)^2 (\bbI + U_{\rm SWAP})
    - b(b+1)\sum_{i=1}^{n}\ket{i,i}\bra{i,i},
\end{equation}
where $U_{\rm SWAP}$ is the swap operator defined by $U_{\rm SWAP}$\( \ket{i,j} = \ket{j, i}  \).
The fourth moment matrix for \( \ket\psi = \calU_W \ket{b\dots b} \) is then \( \sigma^{(2)} = (W^{\otimes 2})\sigma^{(2)}_0 (W^{\otimes 2})^\dag \).
Therefore, given access to \( \sigma^{(2)} \) for \( \ket\psi \), we can form the matrix
\begin{align}
    A &= \frac{1}{b(b+1)}\parentheses{(b+1)^2(\mathbb I + U_{\rm SWAP}) - \sigma^{(2)}} \\
    &=  \sum_{i=1}^n (\ket{w_i}\otimes \ket{w_i})(\bra{w_i}\otimes \bra{w_i}),
\end{align}
where we denote the \( i^{\rm th} \) column vector of \( W \) by \( \ket{w_i} \).

By diagonalizing \( A \) and taking the \( +1 \) eigenvectors, we will find an \( n \)-dimensional subspace spanned by the vectors \( \ket{w_i}\otimes \ket{w_i} \) for \( i=1,\dots,n \).
In particular, a diagonalization algorithm will return the vectors $\ket{\tilde w_i}$, which we can reshape into $n\times n$ normal matrices $C_i$, where
\begin{equation}
    \ket{\tilde w_i} = \sum_{j=1}^n U_{ij}\ket{w_j}\otimes \ket{w_j}, 
    \quad
    C_i = \sum_{j=1}^n U_{ij}\ket{w_j}\bra{w ^{*}_j}
    ,
\end{equation}
for \( i=1,\dots,n \).
Because the eigenspace of $A$ is degenerate, \( U \in \U(n) \) is an arbitrary unitary matrix.
However, because $U$ is unitary, its columns form an orthonormal basis of $\bbC^n$.
This means that the span of the matrices 
$ C_{j} C_{\ell} ^{\dag} $ is a $n^{2}$-dimensional space of matrices, and therefore, up to reordering and phases, there is a unique unitary matrix $V$ that simultaneously diagonalizes all the $C_{j} C_{\ell} ^{\dag} $.
Because $W$ is such a matrix, it follows that by simultaneously diagonalizing all the $C_{j} C_{\ell} ^{\dag} $, we will learn a unitary matrix $V$ that is equal to \( W \) up to a permutation of its columns and global phases applied to the columns (see Supplement for details).
In other words, we have learned the matrix $V = W \Phi P$, where $\Phi$ is some arbitrary diagonal unitary matrix, and $P$ is some arbitrary permutation matrix.
It follows that the state \( \calU_V \ket{b\dots b} \) is the same as \( \calU_W \ket{b\dots b} \) up to an irrelevant global phase, thereby completing the learning algorithm.

We have described the learning algorithm in the special case of a passive Gaussian unitary acting on an initial Fock state \( \ket{b\dots b} \).
The full algorithm for \cref{thm:main-thm-gaussian}, as described in the Supplemental Material, uses both second and fourth moments.
Roughly, second moments are used to learn \( \bm f \) and to rotate (i.e.~block-diagonalize) to blocks where \( f_i = f_{i+1}=\dots \); then we learn the unitaries within each block.

In the End Matter, we describe a few extensions of our algorithm to states beyond those of the form $\calU_S \ket{\bm f}$ as well as a few natural future directions to pursue.

\textit{Comparison to prior learning algorithms.}---%
Ref.~\cite{aaronson2023efficient-tomog}
considered a similar setting to \cref{thm:main-thm-gaussian} in the fermionic case.
Namely, given an unknown fermionic Fock state on $n$ modes that is acted upon by a Gaussian unitary, they devise an efficient state learning algorithm. 
Notably, such a state is a Gaussian fermionic state, and is therefore fully specified by its second moments.
As they point out, their algorithm does not generalize to the case that we consider in this work. 
Ultimately, the reason is because the bosonic states we consider are not Gaussian.
That is, while fermionic Fock states acted upon by a fermionic Gaussian unitary are Gaussian states, bosonic Fock states acted upon by a bosonic Gaussian unitary are \textit{not} Gaussian states.

Ref.~\cite{bittel2025energy-independ} derives a learning algorithm for Gaussian bosonic states whose runtime is independent of the energy.
Again, our algorithm goes beyond Gaussian states.
Nevertheless, it would be very interesting if techniques from Ref.~\cite{bittel2025energy-independ} could be applied to reduce the energy dependence in our algorithm.

Refs.~\cite{mele2024learning-quantu,zhao2024complexity-of-q} consider tomography of non-Gaussian bosonic states.
Notably, however, neither algorithm is efficient in our setting.
Specifically, Ref.~\cite{mele2024learning-quantu} shows that ``$t$-doped'' Gaussian states (the $t$ here should not be confused with the $t$ that we have used throughout this work to denote moments) can be learned in time $\sim n^t$, where the Fock states acted on by Gaussian unitaries that we consider in this work would correspond to $t \sim n$.
Ref.~\cite{zhao2024complexity-of-q} devises a sample-efficient algorithm based on classical shadows that does indeed work for the states that we consider, but the time complexity of the algorithm scales exponentially.
In contrast to Refs.~\cite{mele2024learning-quantu,zhao2024complexity-of-q}, \cref{thm:main-thm-gaussian} provide sample- and time-efficient algorithms.
Importantly, though, there are other classes of non-Gaussian states where our algorithm does not apply but the algorithms in Refs.~\cite{mele2024learning-quantu,zhao2024complexity-of-q} still do apply.
Thus, our results are complementary to
Refs.~\cite{mele2024learning-quantu,zhao2024complexity-of-q}
and provide efficient learning algorithms for different states.

Finally, we note that throughout this work, we have considered the problem of state learning, which is a different problem than unitary learning \cite{Angrisani2025learningunitaries,PhysRevA.81.032324,fanizza,iyer2025mildlyinteractingfermionicunitariesefficiently,austin2025efficientlylearningfermionicunitaries}.


\textit{Conclusion.---}%
In this work, we have considered moments' role in characterizing and learning bosonic quantum states.
Using the moments, we find many quantities that are invariant under Gaussian unitaries.
Our work reveals that Fock states are fully specified by their fourth moments, and we derive an explicit efficient algorithm to learn an unknown Fock state that has been acted upon by an unknown Gaussian unitary, thereby resolving an open problem considered in Ref.~\cite{aaronson2023efficient-tomog}.

There are a number of interesting future directions.
Firstly, for a fixed $t$ (such as $t=6$), one can attempt to find algorithms to perform state tomography on \Gt states.
We expect that the moment methods that we have developed in this work can be extended much more generally to the setting of \Gt states.
In particular, for constant $t$, 
given the promise that a pure state is the unique ground state of a \emph{gapped} degree $t$ Hamiltonian, then the sample complexity for learning that state is automatically $\poly{n}$ by using an argument similar to Ref.~\cite[Cor.~4.3]{yu2023learningmarginalssuffices} to show that knowing the correlation matrices to inverse polynomial precision uniquely specifies the state up to inverse polynomial precision.
We conjecture that one can further always find time efficient learning algorithms for such gapped \Gt states, perhaps by applying the techniques from \cite{rouze2024learning-quantu} to the bosonic setting.
Additionally, the learning problem could potentially be made easier by assuming that the unknown state is of the form $\calU_S \ket{\psi_0}$ for some \textit{known} $\psi_0$, so that the task is only to learn $S$; see the End Matter. 
Another natural open question is addressing \textit{lower bounds} to learning \Gt states.
We discuss this in the End Matter.

Secondly, one may expect that the symplectic invariants described in this work can be proven to fully characterize all invariants of the state, thereby generalizing Williamson's theorem to specify a canonical form of \Gt states.
We suspect that the symplectic invariants could find use in resource theories where Gaussian operations are considered ``free'' \cite{lami2018gaussian-quantu}.

Thirdly, generalizing the learning algorithms and invariants to Gaussian channels, rather than only Gaussian unitaries, is an exciting extension.

Finally, the definition and many properties of \Gt states can be defined for fermionic Gaussian states as well. 
It would be interesting to generate a spectrum of invariants from the fermionic moments, and to consider using our moment techniques to learn non-Gaussian fermionic states.

\begin{acknowledgments}
\textit{Acknowledgments.}---%
We thank Scott Aaronson, Sabee Grewal, Antonio Mele, and Twesh Upadhyaya for stimulating discussions.
J.T.I.~thanks the Joint Quantum Institute at the University of Maryland for support
through a JQI fellowship. J.T.I.~and A.V.G.~acknowledge support from the U.S.~Department of Energy, Office of Science, Accelerated Research in Quantum Computing, Fundamental Algorithmic Research toward Quantum Utility (FAR-Qu). J.T.I.~and A.V.G.~were also supported in part by DARPA SAVaNT ADVENT, ARL (W911NF-24-2-0107), ONR MURI, DoE ASCR Quantum Testbed Pathfinder program (awards No.~DE-SC0019040 and No.~DE-SC0024220), NSF QLCI (award No.~OMA-2120757), NSF STAQ program, AFOSR MURI,  and NQVL:QSTD:Pilot:FTL. J.T.I.~and A.V.G.~also acknowledge support from the U.S.~Department of Energy, Office of Science, National Quantum Information Science Research Centers, Quantum Systems Accelerator (QSA). 
Y.-X.W.~acknowledges support from a QuICS Hartree Postdoctoral Fellowship. 
C.O.~was supported by the National Research Foundation of Korea Grants (No. RS-2024-00431768 and No. RS-2025-00515456) funded by the Korean government (Ministry of Science and ICT~(MSIT)) and the Institute of Information \& Communications Technology Planning \& Evaluation (IITP) Grants funded by the Korea government (MSIT) (No. IITP-2025-RS-2025-02283189 and IITP-2025-RS-2025-02263264).
I.D.~was supported in part by the AFOSR under grants FA9550-21-1-0392 and FA9550-24-1-0089, as well as by a Gerald J. Lieberman Fellowship.
B.F. and S.G. acknowledge support from AFOSR (FA9550-21-1-0008). This material is based upon work partially supported by the National Science Foundation under Grant CCF-2044923 (CAREER), by the U.S. Department of Energy, Office of Science,
National Quantum Information Science Research Centers (Q-NEXT) and by the DOE QuantISED grant DE-SC0020360.
\end{acknowledgments}


\bibliographystyle{apsrev4-2}
\bibliography{references}



\appendix 
\onecolumngrid

\vspace{2em}
\centerline{\bf \Large End Matter}
\vspace{2em}

\twocolumngrid

\section{\texorpdfstring{\Gt}{Gt} states}

In the main text, we noted that
a \Gt state is fully specified by its first $t$ moments.
We now provide more details.

We begin with a mixed \Gt state $\rho$.
Recall that a mixed \Gt state is a thermal state of a degree $\leq t$ Hamiltonian.
Let $\hat M_1,\hat M_2,\dots$ be all the moment operators up to degree $t$, and let $M_i = \Tr[\rho \hat M_i]$.
Suppose Alice gives Bob the moments $M_i$ for all $i$, and Alice promises Bob that those moments came from a \Gt state.
Note that Bob has no access to or knowledge of $\rho$ besides what Alice told him.
Because of the promise, Bob knows that $\rho$ is a Gibbs state of a degree $\leq t$ Hamiltonian.
Using Ref.~\cite{jaynes1957information-the-II}, it follows that $\rho$ is the maximal entropy state subject to constraints on its first $t$ moments. Thus, Bob in principle has enough information to completely reconstruct the state, as he can then perform the following maximization,
\begin{equation}
    \begin{split}
        \rho = \max_\sigma &\parentheses{-\Tr\bargs{\sigma\log\sigma}} \\
        &\text{s.t.~} \forall i\colon~ \Tr[\sigma \hat M_i] = M_i
        .
    \end{split}
\end{equation}

Next, we consider the pure state case.
Recall a pure \Gt state $\psi$ is the unique ground state of a non-degenerate Hamiltonian $H$.
We now want to show that such a state is uniquely specified by its first $t$ moments.
$H$ can be written as $\sum_i c_i \hat M_i$.
Then we have
\begin{salign}
    \bra\psi H \ket\psi
    &= \min_{\ket\phi} \sum_i c_i \bra\phi \hat M_i \ket\phi \\
    \begin{split}
    &= \min_{m_1,m_2,\dots \in \bbC}\sum_i c_i m_i \\
    &\qquad \text{s.t.~there exists a state $\phi$} \\
    &\qquad \text{satisfying }\bra\phi \hat M_i \ket\phi = m_i ~ \forall i 
    \end{split} 
\end{salign}
Let $M_i$ be the minimizing $m_i$ (e.g.~change min to argmin).
Notice that the $\phi$ that satisfies $\bra\phi \hat M_i \ket\phi = M_i$ is uniquely the ground state of $H$, and so $\ket\phi = \ket\psi$.

Now, as before, suppose that Alice gives Bob the numbers $M_i \in \bbC$ and promises that $M_i = \bra\psi \hat M_i \ket\psi$ for a \Gt state $\psi$, but Bob knows nothing else about $\psi$.
With only this information, Bob can in principle reconstruct the state, because from above, he is guaranteed that 
$\ket\psi$ is the unique state that satisfies $M_i = \bra\psi \hat M_i \ket\psi$ for all $i$.

\paragraph*{Discussion.}%
A \Gt state is fully specified by its first $t$ moments in the sense above.
Importantly, we have assumed that the first $t$ moments are known \textit{exactly}. 
If the moments are only known \textit{approximately}, then it is an open and interesting question how accurately one needs to know the first $t$ moments in order to be able to in principle reconstruct the state to a desired fidelity.
Indeed, our learning algorithm in \cref{thm:main-thm-gaussian} is precisely an answer to this question in the case of a restricted class of \G*4 states, namely Fock states acted upon by Gaussian unitaries.
Similarly, Ref.~\cite{mele2024learning-quantu} solves this problem in the case of $t=2$ (i.e.~Gaussian states).

Given the promise that a pure state is the unique ground state of a \emph{gapped} degree $t$ Hamiltonian, then the sample complexity for learning that state is automatically $\poly{n}$ by using an argument similar to Ref.~\cite[Cor.~4.3]{yu2023learningmarginalssuffices} to show that knowing the correlation matrices to inverse polynomial precision uniquely specifies the state up to inverse polynomial precision.
But again, the time complexity of the actually learning the state from the correlation data is unclear.
Our \cref{thm:main-thm-gaussian} is an answer to the latter question for a specific class of Hamiltonians.


\section{Relating states by Gaussian unitaries}

As shown in the main text, given two states $\rho$ and $\sigma$, if their invariants $\lambda(\rho)$ and $\lambda(\sigma)$ are not exactly the same, then they are not related by a Gaussian unitary.
However, it is not immediately obvious whether equality of $\lambda(\rho)$ and $\lambda(\sigma)$ is sufficient to prove that $\rho$ and $\sigma$ be related by a Gaussian unitary.
Specifically, if $\lambda(\rho)$ and $\lambda(\sigma)$ are the same, then invariant functions of $\rho$ and $\sigma$ that are polynomials in the moment tensors must be equal.
It is however not obvious if there always exists a polynomial function in the moment tensors that is able to distinguish two states that are not Gaussian convertible.
In the case of finding invariants under passive Gaussian unitaries in the Hilbert space with a finite Fock cutoff,
polynomials suffice due to the Stone-Weierstrass theorem and the compactness of the unitary group \cite[Prop.~3]{draux2025invariants-in-l}.
In our case, however, the moment tensor entries are not restricted to a compact space, and the symplectic group is not compact.
We leave it as a very interesting open problem whether $\lambda(\rho)$, or even just $\lambda_E(\rho)$, suffices to solve the Gaussian convertibility problem without a finite cutoff in Fock space.
One approach would be to set an energy constraint on the states, thereby effectively making the moment tensor entries compact and the space of active transformations to consider compact.
Further, an interesting question is whether one can truncate $\lambda(\rho)$ to be finite for \Gt states, as the number of moment tensors that one needs to consider is finite.

\section{Extensions of the learning task}

A more general learning task that one can consider is: given \( \ket\psi = \calU_S \ket{\psi_0} \) for some ``initial'' state \( \ket{\psi_0} \), what can we efficiently learn about \( S \) and $\psi_0$?
In this work, we have thus far considered \( \ket{\psi_0} = \ket{\bm f} \).
The algorithm we described for passive Gaussian unitaries \( \calU_W \) acting on \( \ket{b\dots b} \) in fact also succeeds in learning $W$ (modulo a permutation and phases) for any ``GHZ''-type initial state \( \ket{\psi_0} = \sum_{b=0}^\infty c_b \ket{b\dots b} \).
We can see this as follows.
Denote \( 
\overline{b(b+1)} = \sum_b \abs{c_b}^2 b(b+1) \)
and
\( 
\overline{(b+1)^2} = \sum_b \abs{c_b}^2 (b+1)^2 \).
The initial covariance matrix is \( \sigma^{(1)}_0 = 
\overline{b+1}\,
\bbI \), and
the initial fourth moment matrix is \( \sigma^{(2)}_0 = -
\overline{b(b+1)}\sum_i \ket{i,i}\bra{i,i} + \overline{(b+1)^2}
(\bbI + U_{\rm SWAP}) \).
We therefore have that \( \sigma^{(1)} = \sigma^{(1)}_0 \).
Thus, by measuring the second moments, we can learn \( 
\overline{b}\).
Similarly, by measuring the fourth moments, computing the trace, and subtracting off the known \( \overline{b}
\) parts, we can compute \( \overline{b^2}
\).
Then we can use the fourth moment \( \sigma^{(2)} \) to again extract the matrix \( A = W^{\otimes 2} (\sum_i \ket{i,i}\bra{i,i})(W^\dag)^{\otimes 2} \).
Running the remainder of the algorithm on \( A \), we find \( W \) as desired.
Notice, however, that we do not learn the initial state specified by the coefficients \( c_b \).
Indeed,
the GHZ state \( \ket{\pm}\coloneqq \ket{0^n} \pm \ket{1^n} \) has Gaussian degree at least \( n \), because lower moments cannot ``see'' the phase \( \pm \).
Thus, given access to the state $\calU_W \ket{\pm}$, our algorithm can learn $W$ (up to a permutation and phase matrix), but it does not learn the $\pm$ phase.
Instead, once the $W$ is learned, we could apply $\calU_W^\dag$ to the state and then measure a single $n^{\rm th}$ moment in order to learn the $\pm$ phase.

However, our learning algorithm does not work for all initial states \( \ket{\psi_0} \) because it utilizes only information up to fourth moments.
For more general initial states \( \ket{\psi_0} \), it is an interesting question to develop learning algorithms based on the moment matrices.
In particular, given a known \( \ket{\psi_0} \), we can define \( T(\psi_0) \) to be the smallest \( t \) such that \( W \) (resp.~\( S \)) can be learned from the first \( t \) moment matrices of the state \( \calU_W \ket{\psi_0} \) (resp.~\( \calU_S \ket{\psi_0} \) ) for any \( W \) (resp.~\( S \)).
Given this definition, for any \Gt state \( \ket{\psi_0} \), we of course have \( T(\psi_0) \leq t \).
\cref{thm:main-thm-gaussian} proves that for any Fock state \( \ket{\bm f} \), \( T(\bm f) \leq 4 \); more specifically, if \( \bm f \) has all unique elements, then \( T(\bm f) = 2 \), and otherwise \( T(\bm f) = 4 \).
Similarly, any GHZ-like state as defined above has \( T(\text{GHZ}) = 4 \), illustrating that \( T(\ket\psi) \) can be less than the Gaussian degree.

One simple example of a state that requires more than fourth moments to learn the Gaussian unitary is the two mode state \( \ket{\{1,5 \}} \coloneqq \frac{1}{\sqrt 2}\parentheses{\ket{15} + \ket{51}} \).
One can check that \( \sigma^{(1)}_0 \propto \bbI \) and \( \sigma^{(2)}_0 \propto \bbI + U_{\rm SWAP} \).
It follows that \( W \) and \( W^{\otimes 2} \) acting by conjugation have no effect, so that the first four moments contain no information about \( W \) and hence \( T(\set*{1,5}) > 4 \).

This example hints at a general characterization of initial states that completely \emph{hide} a passive Gaussian unitary \( W \) up to \( t^{\rm th} \) moments.
For simplicity, we continue to work within a fixed total boson number subspace so that odd moments can be ignored.
It then follows that the information about \( W \) is \emph{completely hidden} from \( (2t)^{\rm th} \) moments if and only if \( \sigma^{(t)}_0 \) is a sum of permutation matrices on \( (\bbC^n)^{\otimes t} \).
In the \( \ket{\set*{1,5}} \) example, we indeed see that \( \sigma^{(2)}_0 \) is a sum of the two possible permutation matrices on \( (\bbC^n)^{2} \).
The ``if'' direction
comes from \( W^{\otimes t} \) commuting with all permutations.
For the ``only if'', we note that, in order for all \( W \) to be completely hidden, \( \sigma^{(t)}_0 \) must commute with all \( W^{\otimes t} \).
Schur's lemma then implies that \( \sigma_{0}^{\otimes t} \) must be proportional to the identity on the symmetric subspace, and therefore must be a linear combination of permutation operators.
It is an interesting question, with potential cryptographic applications \cite{fefferman2025the-hardness-of}, to construct initial states whose \( T(\psi_0) \) is large in order to most effectively hide unitaries. We leave this to future work.

\section{Sample complexity lower bounds}

\cref{cor:main-result} gives an upper bound on the number of samples required to learn states of the form $\calU_S\ket{\bm f}$ via our efficient algorithm.
A natural question is, can we compute lower bounds on the sample complexity?
We leave finding rigorous lower bounds to future work, but we will briefly argue a lower bound.

Consider the restricted class of states $\calS = \set*{\calU_W \ket{1^n} \mid W \in \U(n)}$.
Two unitaries $W,W'$ give rise to the same state in this set if and only if they are related by a permutation matrix and a diagonal unitary matrix.
Therefore, $\calS$ can be identified with $\calM = \frac{\U(n)}{S_n \ltimes T(\U(n))}$, which is a manifold because $S_n \ltimes T(\U(n))$ is a closed subgroup.
Here, $S_n$ denotes the symmetric group on $n$ elements, and $T(\U(n))$ is the group of diagonal unitary matrices (i.e.,~the maximal torus).
The real dimension of $\calM$ is $n^2 - n$.
Therefore, we expect that at least $\bigOmega*{n^2}$ bits are required to uniquely specify an a point on $\calM$ to some given constant precision.
When measuring from the state, we learn at most $O(n)$ bits of information.
Therefore, we expect that at least $\bigOmega{n}$ samples are required to learn states from $\calS$.

One method to make this argument rigorous is to follow Ref.~\cite[Lem.~S14]{mele2024learning-quantu}.
One must find a finite subset $\calS_\varepsilon \subset \calS$ such that for all $\psi_i,\psi_j\in S_\varepsilon$, $\abs*{\bra{\psi_i}\ket{\psi_i}} \leq 1-\varepsilon$ for some $\varepsilon$. 
Equivalently, one must find a finite subset
$\calM_\varepsilon \subset \calM$
such that for all $U,V \in \calM$, $\abs*{\perm(U^\dag V)} \leq 1-\varepsilon$.
The size of $\abs*{\calS_\varepsilon}$ can be used to put a lower bound on the sample complexity.
Making $\abs*{\calS_\varepsilon}$ as large as possible could help close the gap between a lower bound and our upper bound in \cref{cor:main-result}.
We leave this exciting direction to future work.

\end{document}


\title{Supplemental Material:\texorpdfstring{\\}{}Higher moment theory and learnability of bosonic states}

\author{Joseph~T.~Iosue}
\email{jtiosue@gmail.com}

\author{Yu-Xin~Wang}

\author{Ishaun Datta}

\author{Soumik Ghosh}

\author{Changhun Oh}

\author{Bill Fefferman}

\author{Alexey~V.~Gorshkov}

\maketitle

\tableofcontents



\section{Learning states: main results}

In this section, we review
the main results from the main text that we will be referring to and proving throughout the Supplemental Material.
\cref{thm:main-thm-const-fock,thm:main-thm-arb-fock,thm:main-thm-gaussian} below are more detailed and explicit statements of Theorem 2 of the main text.
In particular, \cref{thm:main-thm-gaussian} is the full statement of Theorem 2 of the main text, and the corresponding algorithm is \cref{alg:main-alg-gaussian}.
\cref{alg:main-alg-gaussian} and \cref{alg:main-alg-arb-fock} both call \cref{alg:main-alg-const-fock} as a subroutine, whose proof of correctness is given in \cref{thm:main-thm-const-fock}.
\cref{cor:nsamples-active} is the full statement of Corollary 3 of the main text.

We note that \cref{thm:main-thm-gaussian} is a learning algorithm for Fock states acted upon by an arbitrary Gaussian unitary, while \cref{thm:main-thm-arb-fock} is more specifically for passive Gaussian unitaries. 
The $n$ dependence is actually better in our bounds in \cref{thm:main-thm-gaussian} than in \cref{thm:main-thm-arb-fock}.
This is an artifact of the $n$ dependence in \cref{eq:block-distance-arb-fock} which is not tight.
Of course, \cref{alg:main-alg-gaussian} from \cref{thm:main-thm-gaussian} can be applied to \cref{alg:main-alg-arb-fock} from \cref{thm:main-thm-arb-fock}, but we still include \cref{thm:main-thm-arb-fock} as it slightly more direct ans because its proof mirrors that of \cref{thm:main-thm-gaussian} but is slightly simpler.

Finally, we emphasize that we very intentionally designed learning \cref{alg:main-alg-const-fock,alg:main-alg-arb-fock,alg:main-alg-gaussian} that use the moment matrices to learn the state as a way of separating the measurement step from the algorithmic step.
There are many methods to measure the moments of a bosonic state, and the best method may depend on the particular physical device that is used.
Nonetheless, we do analyze one method of doing this with heterodyne measurements.
Then, combining \cref{thm:main-thm-const-fock,thm:main-thm-arb-fock,thm:main-thm-gaussian} with measurement protocols in \cref{prop:heterodyne-measure-passive,prop:heterodyne-measure-gaussian}, we arrive at full end-to-end efficient learning algorithms with the guarantees in \cref{cor:nsamples-all1s,cor:nsamples-passive,cor:nsamples-active}.

\medskip

We work with the symplectic group \( \Sp(2n, \bbR) \) defined by the symplectic form \( \Omega = \begin{pmatrix}
    0 & \bbI \\-\bbI & 0
\end{pmatrix} \).
A general Gaussian unitary (modulo displacements, which we will ignore throughout this Supplemental Material because they can be easily learned and accounted for by simply measuring first moments) is specified by a symplectic matrix \( S \in \Sp(2n, \bbR) \), and we denote it by \( \calU_S \) \cite{serafini2017quantum-continu}.
The set of passive (energy-conserving) Gaussian unitaries is then \( K(n) = \Sp(2n, \bbR) \cap \O(2n) \), which is isomorphic to \( \U(n) \) via the isomorphism \( \rho \colon \U(n) \to K(n) \) defined by \cite{serafini2017quantum-continu}
\begin{equation}
    \label{eq:passive-isomorphism}
    \rho(U) = \begin{pmatrix}
        \Re U & -\Im U \\ \Im U & \Re U
    \end{pmatrix}.
\end{equation}
When we consider passive Gaussian unitaries specified by \( W \in \U(n) \), we will denote them by \( \calU_W \), which is understood to mean \( \calU_{\rho(W)} \).

We consider a Fock state \( \ket{\bm f} \), with \( \bm f = (f_1,\dots, f_n) \), on \( n \) modes acted upon by \( \calU_W \) or \( \calU_S \), yielding \( \ket\psi = \calU_W \ket{\bm f} \) or \( \ket\psi = \calU_S \ket{\bm f} \).
Throughout this work, we will ignore all displacements (i.e.,~all first moments are zero) as they can be learned easily by simply measuring first moments.

Given the annihilation \( a_i \) and creation \( a_i^\dag \) operators and the position \( r_i = x_i \) and momentum \( r_{n+i} = p_i \) operators, we define the moment matrices
\begin{salign}
    &\sigma^{(t)}_{i_1,\dots,i_t; j_1,\dots,j_t} = \bra\psi a_{i_1}\dots a_{i_t}a^\dag_{j_1}\dots a^\dag_{j_t}\ket\psi ,\\
    &(\sigma^{(t)}_0)_{i_1,\dots,i_t; j_1,\dots,j_t} = \bra{\bm f} a_{i_1}\dots a_{i_t}a^\dag_{j_1}\dots a^\dag_{j_t}\ket{\bm f} ,\\
    &\Lambda^{(t)}_{i_1,\dots,i_t; j_1,\dots,j_t} = \bra\psi r_{i_1}\dots r_{i_t} r_{j_1}\dots r_{j_t}\ket\psi, \\
    &(\Lambda^{(t)}_0)_{i_1,\dots,i_t; j_1,\dots,j_t} = \bra{\bm f} r_{i_1}\dots r_{i_t} r_{j_1}\dots r_{j_t}\ket{\bm f}.
\end{salign}
Note that these are related to the moments \( \Sigma^{(2t)} \) defined in the main text.
In particular, \( \Lambda^{(t)} \) is contains the same information as \( \Sigma^{(2t)} \), but is a reshaping so that $\Lambda^{(t)}$ is a $(2n)^t \times (2n)^t$ matrix.
The \( \sigma \) matrices can be thought of as submatrices of the \( \Lambda \) matrices that are also orthogonally transformed to convert between the \( (\bm x, \bm p) \) and \( (\bm a, \bm a^\dag) \) bases.

Throughout this work, \( \norm{\cdot} \) refers to the operator norm.
Our main results are \cref{thm:main-thm-const-fock,thm:main-thm-arb-fock,thm:main-thm-gaussian}, which we state now, and \cref{cor:nsamples-active}, which we state and prove in \cref{sec:measurement}.

\begin{theorem}
    \label{thm:main-thm-const-fock}
    Suppose \( \bm f = (b,\dots, b) \) for some nonnegative integer \( b \) and let \( \ket\psi = \calU_W \ket{\bm f} \) for an unknown unitary \( W \in \U(n) \) specifying an arbitrary passive Gaussian unitary.
    If our measurement \( \sigma^{(2)\prime} \) of the moment matrix \( \sigma^{(2)} \) for \( \ket\psi \) satisfies \( \norm*{ \sigma^{(2)\prime} - \sigma^{(2)} } \leq \varepsilon \), then we can efficiently (via \cref{alg:main-alg-const-fock}) find a $V \in \U(n)$ such that
    \begin{equation}
        \norm*{V - W\Phi P} \leq \frac{4\sqrt{5}\varepsilon n}{b(b+1)}
    \end{equation}
    for some (irrelevant) diagonal unitary matrix \( \Phi \) and permutation matrix \( P \).
    In particular,
    \begin{equation}
        \abs*{\bra{b^n} \calU_W^\dag \calU_V \ket{b^n}} \geq
        1 - \frac{4\sqrt{5}\varepsilon n^2 / (b+1)}{1 - 4\sqrt{5}\varepsilon n^2 / (b+1)}
        %
    \end{equation}
    as long as
    \( \varepsilon \leq \frac{b+1}{4\sqrt{5} n^2} \).
\end{theorem}
\begin{proof}
    See \cref{sec:ideal-description-const-fock,sec:noisy-description-const-fock}
\end{proof}

\begin{theorem}
    \label{thm:main-thm-arb-fock}
    Let \( \ket\psi = \calU_W \ket{\bm f} \) for an unknown unitary \( W \in \U(n) \) specifying an arbitrary passive Gaussian unitary and an arbitrary Fock state \( \ket{\bm f} \).
    If our measurements \( \sigma^{(1)\prime},\sigma^{(2)\prime} \) of the moment matrices \( \sigma^{(1)},\sigma^{(2)} \) satisfy \( \norm*{\sigma^{(t)\prime}-\sigma^{(t)}} \leq \varepsilon_t \),
    then we can efficiently (via \cref{alg:main-alg-arb-fock}) find a \( V\in \U(n) \) and \( \bm g \) such that
    \( \norm{V - W \Phi P} \leq \gamma \), with
    \begin{equation}
        \gamma =
        \varepsilon_1 \parentheses{
            32\sqrt5 n^2 (3 f_{\rm max}^2 + 5 f_{\rm max} + 2)
            + 4n
        }
        +
        2\sqrt5 \varepsilon_2 n
    \end{equation}
    for some diagonal unitary matrix \( \Phi \) and a permutation matrix \( P \), and \( f_{\rm max} = \max_i f_i \).
    Specifically, \( \bm g \) is some permutation of \( \bm f \) and \( P \) performs this permutation along with other (irrelevant) permutations within blocks of equal \( g_i \).
    In particular,
    \begin{equation}
        \abs*{\bra{\bm f} \calU_W^\dag \calU_V \ket{\bm g}} \geq 1 - \frac{\gamma f_{\rm max} n}{1 - \gamma f_{\rm max} n}
        %
    \end{equation}
    as long as \( \gamma f_{\rm max} n < 1 \).
\end{theorem}
\begin{proof}
    See \cref{sec:ideal-description-arb-fock,sec:noisy-description-arb-fock}
\end{proof}

We note that \cref{thm:main-thm-gaussian} applied to the passive case actually gives better bounds than our direct proof of \cref{thm:main-thm-arb-fock}. 
The reason can be tracked to the $n$-dependence of \cref{eq:block-distance-arb-fock}, which is not tight.
However, we include \cref{thm:main-thm-arb-fock} anyways because it is easier to explicitly track the constants, and also because its proof mirrors that of \cref{thm:main-thm-gaussian} but is slightly simpler.
We emphasize though that setting $s=0$ in \cref{thm:main-thm-gaussian} gives a better bound than \cref{thm:main-thm-arb-fock}.

\begin{theorem}[Theorem 2 of the main text]
    \label{thm:main-thm-gaussian}
    Let \( \ket\psi = \calU_S \ket{\bm f} \) for an unknown symplectic matrix \( S \in \Sp(2n, \bbR) \) specifying an arbitrary Gaussian unitary and an arbitrary Fock state \( \ket{\bm f} \).
    If our measurements \( \Lambda^{(1)\prime},\Lambda^{(2)\prime} \) of the moment matrices \( \Lambda^{(1)},\Lambda^{(2)} \) satisfy \( \norm*{\Lambda^{(t)\prime} - \Lambda^{(t)}} \leq \varepsilon_t \),
    then we can efficiently (via \cref{alg:main-alg-gaussian}) find a \( Q \in \Sp(2n, \bbR) \) and \( \bm g \) such that
    \( \norm{Q - S \Phi P} \leq \gamma \), where
    \begin{equation}
        \gamma = \bigO{
        \varepsilon_1 n \e^{2s}  f_{\rm max}^2 + \varepsilon_2 n \e^{5s}
        }
    \end{equation}
    for some symplectic matrices \( \Phi \) and \( P \) that implement global phases and mode permutations, \( f_{\rm max} = \max_i f_i \), and \( s \) is the maximum magnitude of squeezing in \( S \) (that is, \( \e^s \) is the largest singular value of \( S \)).
    Specifically, \( \bm g \) is some permutation of \( \bm f \) and \( P \) performs this permutation along with other (irrelevant) permutations within blocks of equal \( g_i \).
    In particular,
    \begin{equation}
        \abs*{\bra{\bm f}\calU_S^\dag \calU_Q \ket{\bm g}} \geq
        1
        -
        \bigO{\gamma \e^s n f_{\rm max}}.
    \end{equation}
\end{theorem}
\begin{proof}
    See \cref{sec:ideal-description-gaussian,sec:noisy-description-gaussian}
\end{proof}

The remainder of this Supplemental Material is organized as follows.
In \cref{sec:ideal-description-const-fock}, we describe \cref{alg:main-alg-const-fock}, which proves \cref{thm:main-thm-const-fock} in the ideal case when the moments are known exactly.
In \cref{sec:noisy-description-const-fock}, we prove \cref{thm:main-thm-const-fock} by analyzing the case when the moments are known only to norm precision.
In \cref{sec:ideal-description-arb-fock}, we describe \cref{alg:main-alg-arb-fock}, which proves \cref{thm:main-thm-arb-fock} in the ideal case when the moments are known exactly.
In \cref{sec:noisy-description-arb-fock}, we prove \cref{thm:main-thm-arb-fock} by analyzing the case when the moments are known only to norm precision.
Note that the proof of \cref{thm:main-thm-arb-fock} uses \cref{thm:main-thm-const-fock}.
In \cref{sec:ideal-description-gaussian}, we describe \cref{alg:main-alg-gaussian}, which proves \cref{thm:main-thm-gaussian} in the ideal case when the moments are known exactly.
In \cref{sec:noisy-description-gaussian}, we prove \cref{thm:main-thm-gaussian} by analyzing the case when the moments are known only to norm precision.
Note that the proof of \cref{thm:main-thm-gaussian} uses \cref{thm:main-thm-const-fock}.

Finally, note that,
given $\poly{n}$ copies of a quantum state, the moments can be measured to $1/\poly{n}$ precision. 
Thus, for \cref{thm:main-thm-const-fock,thm:main-thm-arb-fock,thm:main-thm-gaussian}, we assume that that the moments are known to a given precision and show how the state can be accurately reconstructed from these moments. 
In \cref{sec:measurement}, using \cref{thm:main-thm-const-fock,thm:main-thm-arb-fock,thm:main-thm-gaussian} and various probability bounds with a median of means heterodyne estimator, we analyze the end-to-end learning algorithm---that is, we derive explicit bounds on the number of heterodyne measurements from the state that are needed in order to learn the state to a desired fidelity (see \cref{cor:nsamples-all1s,cor:nsamples-passive,cor:nsamples-active}).
In \cref{sec:numerics}, we numerically simulate the learning algorithm for states $\calU_W\ket{1^n}$ for random unknown $W \in \U(n)$.
This complements our rigorous bounds on the number of required measurements with numerical fits showing considerably fewer expected number of required measurements.


\section{Description of the algorithm in Theorem S1}
\label{sec:ideal-description-const-fock}

In this section, we prove \cref{thm:main-thm-const-fock} in the error-free (\( \varepsilon = 0 \)) case by describing the full algorithm, which we summarize in \cref{alg:main-alg-const-fock}.
This algorithm was concisely described in the main text, but we expand upon it in this section, as well as set the relevant notation for the proofs in the remainder of the Supplemental Material.
In the next section, \cref{sec:noisy-description-const-fock}, we prove the full theorem.

\begin{algorithm}
    \caption{Algorithm in \cref{thm:main-thm-const-fock}}\label{alg:main-alg-const-fock}
    \begin{algorithmic}[1]

        \Procedure{findV}{$\sigma^{(2)}, b$}

        \If{\( b = 0 \)}
        \State \Return \( \bbI \)
        \EndIf

        \State \( A \gets \frac{1}{b(b+1)} \parentheses{(b+1)^2(\bbI \otimes \bbI + \text{SWAP}) - \sigma^{(2)}} \)

        \State \( V \gets \) the \( n\times n \) zero matrix
        \State \( j \gets 1 \)

        \For {\( i=1,\dots,n \)}
        \State \( \ket{\tilde w_i} \gets \) eigenvector of \( A \) corresponding to the \( i^{\rm th} \) largest eigenvalue
        \State $C_i \gets$ reshape $\ket{\tilde w_i}$ into a $n\times n$ matrix
        \ignore{
        \For{(eigenvalue, eigenspace) \( (\lambda, \text{span}\set*{v_1,\dots, v_d}) \) in eigendecomposition of \( C_i \)}
        \If{$d = 1$}
        \State Set the \( j^{\rm th} \) column of \( V \) to be \( \ket{v_1} \)
        \If{rank\((V) = j\)}
        \If{\( j = n \)}
        \State \Return \( V \)
        \EndIf
        \State \( j \gets j + 1 \)
        \Else
        \State Set the \( j^{\rm th} \) column of \( V \) to be \( 0 \)
        \EndIf
        \EndIf
        \EndFor
        }
        
        \EndFor

        \State Find a unitary $V$ the simultaneously diagonalizes $C_1,\dots,C_n$ (see \cref{sec:ideal-description-const-fock})
        \State \Return $V$

        \EndProcedure

    \end{algorithmic}

\end{algorithm}

Thus, throughout this section, we assume that we know \( \sigma^{(2)} \) for \( \calU_W \ket{b^n} \) perfectly, where we denote the \( n \)-mode Fock state \( \ket{b\dots b} \) by \( \ket{b^n} \). We show that, by using \( \sigma^{(2)} \), we can find a unitary \( V \in \U(n) \) such that \( V = W \Phi P \), where \( \Phi \) is an arbitrary diagonal unitary matrix and \( P \) is an arbitrary permutation matrix.
From this, it then follows that
\begin{equation}
    \abs*{\bra{b^n}\calU_W^\dag \calU_V \ket{b^n}} = \abs*{\bra{b^n}\calU_\Phi \calU_P \ket{b^n}} = \abs*{\det\Phi} = 1,
\end{equation}
as desired.

The initial correlators are
\begin{equation}
    (\sigma^{(2)}_0)_{ij;kl}
    = \bra{b^n} a_i a_j a_k^\dag a_l^\dag \ket{b^n}
    = \begin{cases}
        (b+1)^2    & \text{if } \{i,j \} = \{k,l \} \text{ and } i\neq j \\
        (b+1)(b+2) & \text{if } i = j = k = l                            \\
        0          & \text{otherwise}
    \end{cases} .
\end{equation}
It therefore follows that
\begin{salign}
    \sigma_0^{(2)}
    &= (b+1)(b+2)\sum_i \ket{i,i}\bra{i,i} + (b+1)^2 \sum_{i\neq j}(\ket{i,j}\bra{i,j}+\ket{i,j}\bra{j,i}) \\
    &= -b(b+1)\sum_i \ket{i,i}\bra{i,i} + (b+1)^2\sum_{i, j}(\ket{i,j}\bra{i,j}+\ket{i,j}\bra{j,i}) \\
    &= -b(b+1)\sum_i \ket{i,i}\bra{i,i} + (b+1)^2(\mathbb I + \text{SWAP}),
\end{salign}
and thus
\begin{equation}
    \sigma^{(2)}
    = (W\otimes W)\sigma_0^{(2)} (W\otimes W)^\dag
    = -b(b+1)\sum_i (W\otimes W)\ket{i,i}\bra{i,i}(W^\dag \otimes W^\dag) + (b+1)^2(\mathbb I + \text{SWAP}).
\end{equation}

Denote the $i^{\rm th}$ column vector of $W$ by $\ket{w_i}$. We see that if we can measure the moments $\sigma^{(2)}$, then we can compute the matrix
\begin{equation}
    A = \frac{1}{b(b+1)}\parentheses{(b+1)^2(\mathbb I + \text{SWAP}) - \sigma^{(2)}} =  \sum_{i=1}^n (\ket{w_i}\otimes \ket{w_i})(\bra{w_i}\otimes \bra{w_i}).
\end{equation}
Therefore, given $A$, we want to determine each $w_i$. Note that, because the whole problem is permutation symmetric, we do not care about the ordering of the $w_i$'s.
By diagonalizing $A$ and taking the $+1$ eigenvectors, we will find vectors $\{\ket{\tilde w_i} \mid i=1,\dots,n\}$, where
\begin{equation}
    \ket{\tilde w_i} = \sum_{j=1}^n U_{ij}\ket{w_j}\otimes \ket{w_j}
    =
    \sum_{j=1}^n \abs*{U_{ij}}e^{i\phi_{ij}}\ket{w_j}\otimes \ket{w_j}
\end{equation}
for some unitary matrix $U$ that we do not know. From the Schmidt decomposition \cite{nielsen2010quantum-computa}, the $\abs*{U_{ij}}$ are unique up to reordering. Thus, we have learned $\ket{w_i}$ for each $i$ up to phase.

We note, however, that there is a slight subtlety here.
The Schmidt basis, which is what we will learn for $\ket{w_i}$, is only unique if there is no degeneracy in the $\abs{U_{ij}}$ coefficients.
Specifically,
we can turn $\ket{\tilde w_i}$ into a normal matrix $C_i$ by flipping the second ket to a bra, therefore arriving at $C_i = \sum_j U_{ij} \ket{w_j}\bra{w ^{*}_j}$, where $\ket{w ^{*}_j}$ denotes the basis vectors that are complex conjugate to $\ket{w_j}$.
By taking the eigensystem of $C_{j} C_{\ell} ^{\dag} = \sum_j U_{ij} U_{\ell j} ^{*} \ket{w_j}\bra{w_j}$, we are only guaranteed to learn the true $\ket{w_j}$ if the eigenvalues of $C_{j} C_{\ell} ^{\dag} $ (which are $U_{ij} U_{\ell j} ^{*} $) are non-degenerate. 
But recall that we have no control over the matrix $U$.
To guarantee that we learn the correct $\ket{w_j}$, we proceed as follows.

First we will describe an easy method that succeeds with probability $1$ and is thus a good choice in practice. In the next paragraph, we will describe a deterministic method that always succeeds.
Randomly draw a unitary $Y$ from the Haar measure on $\U(n)$, and create the matrices $D_i = \sum_{j,\ell} Y_{ij}C_j C_{\ell} ^{\dag} Y_{i\ell} ^{*}$.
It follows that $D_i = \sum_{k}| (YU)_{ik} |^{2} \ket{w_k}\bra{w_k}$.
Because $Y$ is Haar random, $YU$ is also Haar random.
Therefore, with probability $1$
$D_i$ will have non-degenerate eigenvalues.
Thus, by diagonalizing any of the $D_i$, we will learn the true $\ket{w_k}$ up to phase.

A deterministic method to learn the $\ket{w_j}$ regardless of $U$ is as follows. 
Notice that because $U$ is unitary and the $\ket{w_j}$ form an orthonormal basis, up to permutations and phases there is a unique unitary $V$ that simultaneously diagonalizes all of the $C_{j} C_{\ell} ^{\dag}$. $W$ is such a unitary, and therefore by simultaneously diagonalizing all of the $C_{j} C_{\ell} ^{\dag}$, we will find a $V$ satisfying $V = W \Phi P$, where $\Phi$ is some arbitrary diagonal unitary matrix and $P$ is some arbitrary permutation matrix.
Thus, we just need to simultaneously diagonalize all of the $C_{j} C_{\ell} ^{\dag}$.
We now describe a method to do this.
First, diagonalize $C_1 C_{1} ^{\dag}$ as $C_1  C_{1} ^{\dag} = V^{(1)} D^{(1)} V^{(1)\dag}$. 
Then let $\tilde C_2 \tilde C_2 ^{\dag} = V^{(1)\dag} C_2 C_{2} ^{\dag} V^{(1)}$.
Because $C_1 C_{1} ^{\dag}$ and $C_2 C_{2} ^{\dag} $ commute, $\tilde C_2  \tilde C_2 ^{\dag}$ will be block diagonal, $\tilde C_2 = \bigoplus_j B^{(2)}_j$.
Diagonalize each block as $B^{(2)}_j = V^{(2)}_j D^{(2)}_j V^{(2)\dag}_j$.
Set $V^{(2)} = \bigoplus_j V^{(2)}_j$.
By construction, $V^{(1)}V^{(2)}$ diagonalizes both $C_1 C_{1} ^{\dag}$ and $C_2 C_{2} ^{\dag}$.
Next, let $\tilde C_3 \tilde C_3 ^{\dag} = (V^{(1)} V^{(2)})^\dag C_3 C_3 ^{\dag} V^{(1)} V^{(2)}$. Again $\tilde C_3 \tilde C_3 ^{\dag}$ is block diagonal.
Diagonalize each block as $B^{(3)}_j = V^{(3)}_j D^{(3)}_j V^{(3)\dag}_j$ and
set $V^{(3)} = \bigoplus_j V^{(3)}_j$.
Proceeding in this manner, we set $V = V^{(1)}V^{(2)}\dots V^{(n)}$.

If there are any degenerate eigen-subspace shared by $C_i C_{i} ^{\dag}$ for all $i$, we then repeat the similar procedure for all $C_i C_{j} ^{\dag}$ with $i \ne j$, which must break degeneracy for all the $\ket{w_k}$ eigenkets of the $C_i C_{j} ^{\dag}$ matrices. This is because if the $C_i C_{j} ^{\dag}$ matrices share a degenerate set of eigenbases for all $i$ and $j$ indices, suppose two of the eigenbasis states are $\ket{w_k}$ and $\ket{w_\ell}$, then this would imply that $U_{ik} U_{j k} ^{*} = U_{i \ell} U_{j \ell} ^{*} $ for all $i$ and $j$, which would  require that $U_{ik} =U_{i\ell} $ for all $i$. However, this would lead to a contradiction with the assumption that $U$ is unitary. Hence, we have shown that above algorithm always generates a unitary transformation $V$ that is unique up to permutation and phase shifts of the basis vectors $\ket{w_k}$.

\ignore{
Let $S = \set*{}$.
Diagonalize the matrix $C_1$. For every unique eigenvalue, add the corresponding eigenvector to $S$.
Let the degeneracy be $d = n - \abs{S}$.
If $d = 1$, then we are done, and $S = \set{\e^{\i \phi_i}\ket{w_i} \mid i=1,\dots n}$ for some unknown and irrelevant phases $\phi_i$.
If $d \geq 2$, then we move on to $C_i$ for $i \geq 2$.
We can without loss of generality then assume that $U_{11} =\dots = U_{1d}$.
By unitarity of $U$, it cannot be that $U_{i1}=\dots=U_{id}$ for all $i=1,\dots, n$.
Indeed, this means that for any degenerate (of any size) in any of the $C_i$, there is at least one $j$ such that $C_j$ does not have that degenerate subspace. 
As an example, suppose that $C_1$ has two two-fold degenerate spaces $U_{11}=U_{12}$ and $U_{13} = U_{14}$ an the rest are nondegenerate.
\note{finish}
Therefore, we can find all of the $\ket{w_j}$ by iterating through all of the $C_i$, diagonalizing $C_i$, and adding all eigenvectors $\ket v$ corresponding to unique eigenvalues to $S$ as long as $v$ is not already in the span of $S$.

\note{Hm the above may not work. Maybe need to do this instead:}
\note{
Randomly draw a unitary $O$ from the Haar measure on $\U(n)$, and create the matrices $D_i = \sum_j O_{ij}C_j$.
It follows that $D_i = \sum_{k}(OU)_{ik} \ket{w_k}\bra{w_k}$.
Because $O$ is Haar random, $OU$ is also Haar random.
Therefore, with probability $1$ \note{given machine precision, there is technically some finite but tiny probability less than $1$}
$D_i$ will have non-degenerate eigenvalues.
Thus, by diagonalizing any of the $D_i$, we will learn the true $\ket{w_k}$ up to phase.
\note{Do we need to put this into the error analysis?}
\note{Do we need to add this Haar random step to \cref{alg:main-alg-const-fock}?}
\note{Is there a better way to do this without introducing any randomness?}
}

\note*{
This problem is equivalent to the following. 
Let $\pi_i = \ket{w_i}\bra{w_i}$ for $i=1,\dots, n$.
Given a bitstring $\bm b \in \{0,1\}^n$, define $\Pi_{\bm b} = \sum_{i=1}^n b_i \pi_i$.
Suppose we are given a collection of bitstrings $\bm b^{(1)},\dots, \bm b^{(D)}$ and their corresponding projectors $\Pi_{\bm b^{(i)}}$.
Via our argument above, we are promised that for any pair $i\neq j$, there is at least one $k$ for which $b^{(k)}_i \neq b^{(k)}_j$.
Our task is to learn each $\pi_i$.

A few useful facts. First, rank$(\Pi_{\bm b}) = \abs*{\bm b}$, where $\abs{\cdot}$ denotes the Hamming weight. 
Next, multiplication of projectors corresponds to taking the AND of the bitstrings:
$\Pi_{\bm b}\Pi_{\bm c} = \Pi_{\bm b \land \bm c}$, and thus the projectors commute.
Next, the XOR of bitstrings corresponds to
$(\Pi_{\bm a} - \Pi_{\bm b})^2 = \Pi_{\bm a \oplus \bm b}$.

Given the projectors, we can easily learn the corresponding bitstrings (up to a global permutation) (we show this below).
Now, we simply need to see how we can take $\bm b^{(1)}, \dots, \bm b^{(D)}$ and produce the $n$ different Hamming weight $1$ bitstrings using the AND and XOR operations.
Given our argument above about there being bitstrings with different values for every pair, it follows (make this rigorous) that $\text{span}_{\bbF_2}(\bm b^{(1)}, \dots, \bm b^{(D)}) = \bbF_2^n$. Thus, using just the XOR operation (ie addition in $\bbF_2$), we can generate all $n$ Hamming weight $1$ bitstrings.
This then gives us all the $\pi_j$.

So now all we need to show is that we can learn the bitstrings up to a global permutation.
In particular, notice that given the projectors, we can compute the rank of products of projectors.
Thus, the task becomes the following:
find a set of bitstrings $\bm b^{(1)}, \dots, \bm b^{(D)}$ satisfying constraints on $\abs{\bigwedge_{j} \bm b^{(i_j)}}$.
In other words, we have the ability to compute the Hamming weight of any AND of any combination of the bitstrings, and we need to find any set of bitstrings that has those Hamming weights.
We can do this algorithmically.
Let's show an example. Suppose that $\abs{\bm b^{(1)}} = 4$. Then we set $\bm b^{(1)} = (1,1,1,1,0,0,\dots)$.
Suppose that $\abs{\bm b^{(2)}} = 3$ and $\abs{\bm b^{(1)}\land \bm b^{(2)}} = 2$. Then we can set $\bm b^{(2)} = (1,1,0,0,1,0,0,\dots)$.
Suppose that $\abs{\bm b^{(3)}} = 4$, $\abs{\bm b^{(1)}\land \bm b^{(3)}} = 3$, $\abs{\bm b^{(2)}\land \bm b^{(3)}} = 2$. And I think you also need $\abs{\bm b^{(1)}\land \bm b^{(2)}\land \bm b^{(3)}}$.
And so on.

There's clearly a way to solve this problem by looking at all of these ``intersections''. But as is it seems very inefficient. Can we do this more efficiently? Maybe we just need to have some guarantee's on the structure of the bitstrings. These guarantees would come from the properties of $U$.

Notice that $(\bm a \oplus \bm b) \land \bm a$ is the bitstring that is $1$ everywhere that $\bm a$ is $1$ and $\bm b$ is $0$.
Can we use this?
I might be able to use this to keep the Hamming weights small. Because when there are intersections, this shrinks the Hamming weight. We just need to make sure everything stays linearly independent.

Ah I see, we do something like this.
WE set $b^{(1)}$ arbitrarily.
Then if the hammingweight of the intersection of one and two is nothing, then we set $b^{(2)}$ arbitrarily.
If the hamming weight of the intersection is everything and their ranks are equal, then we just ignore.
Otherwise, we convert $\Pi_{\bm b^{(2)}} \to \Pi_{\bm b^{(1)}\land \bm b^{(2)}}$ and set $b^{(2)}$ to be the overlap with $b^{(1)}$.
Then we add a $b^{(2')}$ that is the original $b^{(2)}$ xor with the new $b^{(2)}$ (something like this?).
Etc.
}
}

We have therefore learned the matrix $W$ up to permutation of the columns and up to global phases in each column. In other words, we have learned the matrix $V = W \Phi P$, where $\Phi$ is some arbitrary diagonal unitary matrix, and $P$ is some arbitrary permutation matrix.

In summary, if we have access to \( \sigma^{(2)} \) for the state $\calU_W\ket{b^n}$ for some unknown $W$, we can create and diagonalize the matrix $A$, and store the $+1$ eigenvectors. We perform the Schmidt decomposition on each of these eigenvectors. The resulting vectors as columns give us a matrix $V$ such that $\abs*{\bra{b^n}\calU_W^\dag \calU_V \ket{b^n}} = 1$.
This therefore gives \cref{alg:main-alg-const-fock}.

\section{Proof of Theorem S1}
\label{sec:noisy-description-const-fock}

In this section, we prove \cref{thm:main-thm-const-fock} by analyzing the effect that an error in our knowledge of the moments has on \cref{sec:ideal-description-const-fock,alg:main-alg-const-fock}.
We use the same notation as in the description of the algorithm in \cref{sec:ideal-description-const-fock}, and thus all definitions carry over.
We prove \cref{thm:main-thm-const-fock} by proving that, when given \( \sigma^{(2)\prime} = \sigma^{(2)} + \varepsilon E \) with \( \norm{E} \leq 1 \) instead of \( \sigma^{(2)} \), \cref{alg:main-alg-const-fock} will yield a \( V \) satisfying the statement of \cref{thm:main-thm-const-fock}.
Then when we compute $A$, we actually compute $A' = A - \frac{\varepsilon}{b(b+1)}E$.
Throughout the remainder of this section, we will set \( \varepsilon \to b(b+1)\varepsilon/2 \); we will put back the factor at the end.
The eigenvalues of $A'$ will be within $\sim \varepsilon \lVert E \rVert$ of the eigenvalues of $A$
\cite[Thm.~VI.5.1,~VII.4.1]{bhatia1997matrix-analysis}.
Assuming $\varepsilon$ is small, we take the largest $n$ eigenvalues/eigenvectors of $A'$, which will have eigenvalues $\geq 1 - \varepsilon$. The other $n^2-n$ eigenvectors will have eigenvalues $\leq \varepsilon$.

Notice that $A$ is a projector onto the desired eigenspace (\emph{i.e.}~the $\ket{\tilde w_i}$'s).
For $i = 1,\dots, n$,
let $\ket{\tilde w_i^\prime}$ be the largest $n$ eigenvectors of $A'$. We are interested in the distance between each $\ket{\tilde w_i^\prime}$ and the desired eigenspace, which is the image of the projector $A$. Thus, we are interested in (\emph{i.e.}~we will use later)
\begin{salign}
    \lVert \tilde w_i' - A \tilde w_i' \rVert^2
    &= \lVert \tilde w_i' - A' \tilde w_i' - \frac{\varepsilon}{2}E\tilde w_i' \rVert^2 \\
    &\leq \lVert \tilde w_i' - A' \tilde w_i'\rVert^2 +\frac{\varepsilon^2}{4}\lVert E\tilde w_i' \rVert^2 \\
    &\leq \lVert \tilde w_i' - A' \tilde w_i'\rVert^2 +\frac{\varepsilon^2}{4} \\
    &\leq \lVert \tilde w_i' -(1-\varepsilon) \tilde w_i'\rVert^2 +\frac{\varepsilon^2}{4} \\
    &= \frac{5\varepsilon^2}{4}.
\end{salign}
This implies that
\begin{equation}
    \label{eq:projector-distance}
    \implies \sum_i \lVert \tilde w_i^\prime - A \tilde w_i^\prime\rVert^2 \leq \frac{5 \varepsilon^2 n}{4}.
\end{equation}
We consider the Schmidt decomposition of $\ket{\tilde w_i^\prime}$,
\begin{equation}
    \begin{aligned}
        \ket{\tilde w_i^\prime} = \sum_j U_{ij}' \ket{w_j'} & \otimes \ket{w_j''}, \quad \text{where}           \\
                                                            & \ket{w_j'} = c_j' \ket{w_j} + s_j' \ket{v_j'},     \\
                                                            & \ket{w_j''} = c_j'' \ket{w_j} + s_j'' \ket{v_j''}, \\
                                                            & c_j' = \cos\theta_j', s_j' = \sin\theta_j',        \\
                                                            & c_j'' = \cos\theta_j'', s_j'' = \sin\theta_j'',    \\
                                                            & \bra{w_j}\ket{v_j'} = \bra{w_j}\ket{v_j''} = 0.
    \end{aligned}
\end{equation}
We want to show that $s^2 \coloneqq \max_j \max(s_j^{\prime 2}, s_j^{\prime\prime 2})$ is small---this will tell us that the vectors that we find, $\ket{w_i^\prime}$, are close to the ones we would find in the noiseless case, $\ket{w_i}$. Beginning with \cref{eq:projector-distance}, we compute
\begin{salign}
    \frac{5 \varepsilon^2 n}{4}
    &\geq \sum_i
    \lVert \tilde w_i^\prime - A \tilde w_i^\prime\rVert^2 \\
    &= \sum_i \bra{\tilde w_i^\prime - A \tilde w_i^\prime}\ket{\tilde w_i^\prime - A \tilde w_i^\prime} \\
    &= \sum_i \bra{\tilde w_i'}(\bbI - A)^2 \ket{\tilde w_i'}\\
    &= n - \sum_i\bra{\tilde w_i^\prime} A \ket{\tilde w_i^\prime} \\
    \begin{split}
        &= n - \sum_{i,j,k} \bar U'_{ij} U'_{ik}
        \left(c_j' c_j''\bra{w_j w_j} + s_j' s_j'' \bra{v_j'v_j''} + c_j's_j'' \bra{w_jv_j''} + c_j''s_j' \bra{v_j'w_j}\right)
        \\
        &\qquad \times A \left(c_k' c_k'' \ket{w_k w_k} + s_k' s_k'' \ket{v_k'v_k''} + c_k's_k'' \ket{w_kv_k''} + c_k''s_k' \ket{v_k'w_k}\right)
    \end{split} \\
    \begin{split}
        &= n - \sum_{j}
        \left(c_j' c_j''\bra{w_j w_j} + s_j' s_j'' \bra{v_j'v_j''} + c_j's_j'' \bra{w_jv_j''} + c_j''s_j' \bra{v_j'w_j}\right)
        \\
        &\qquad \times A \left(c_j'c_j''\ket{w_j w_j} + s_j' s_j'' \ket{v_j'v_j''} + c_j's_j'' \ket{w_jv_j''} + c_j''s_j' \ket{v_j'w_j}\right)
    \end{split}\\
    \begin{split}
        &= n - \sum_{j}
        \left(c_j' c_j''\bra{w_j w_j} + s_j' s_j'' \bra{v_j'v_j''} + c_j's_j'' \bra{w_jv_j''} + c_j''s_j' \bra{v_j'w_j}\right)
        \\
        &\qquad \times A^2 \left(c_j' c_j''\ket{w_j w_j} + s_j' s_j'' \ket{v_j'v_j''} + c_j's_j'' \ket{w_jv_j''} + c_j''s_j' \ket{v_j'w_j}\right)
    \end{split}\\
    \begin{split}
        &= n - \sum_{j}
        \left(c_j' c_j'' \bra{w_j w_j} + s_j' s_j'' \bra{v_j'v_j''} A\right)
        \\
        &\qquad \times\left(c_j' c_j''\ket{w_j w_j} + s_j' s_j'' A \ket{v_j'v_j''} \right)
    \end{split} \\
    &= n - \sum_{j} \left(c_j^{\prime 2} c_j^{\prime\prime 2} + s_j^{\prime 2} s_j^{\prime\prime 2}\bra{v_j'v_j''}A\ket{v_j'v_j''} + 2c_j' c_j'' s_j's_j'' \text{Re}\bra{v_j'v_j''}A\ket{w_jw_j} \right) \\
    &= n - \sum_{j} \left(c_j^{\prime 2} c_j^{\prime\prime 2} + s_j^{\prime 2} s_j^{\prime\prime 2} \bra{v_j'v_j''}A\ket{v_j'v_j''} \right) \\
    &\geq n - \sum_{j} \left(c_j^{\prime 2} c_j^{\prime\prime 2} + s_j^{\prime 2} s_j^{\prime\prime 2}  \right) \\
    &= n - \sum_{j} \left(
    1 - s_j^{\prime 2} - s_j^{\prime \prime 2} + s_j^{\prime 2}s_j^{\prime\prime 2}
    + s_j^{\prime 2} s_j^{\prime\prime 2}  \right) \\
    &= \sum_{j} \left(s_j^{\prime 2} + s_j^{\prime \prime 2} - 2 s_j^{\prime 2}s_j^{\prime\prime 2}  \right) \\
    &\geq \sum_{j} \left(\max(s_j^{\prime 2}, s_j^{\prime \prime 2}) - 2 \max(s_j^{\prime 4}, s_j^{\prime \prime 4})  \right) \\
    &\geq \max_{j} \left(\max(s_j^{\prime 2}, s_j^{\prime \prime 2}) - 2 \max(s_j^{\prime 4}, s_j^{\prime \prime 4})  \right) \\
    &\geq s^2 - 2 s^4.
\end{salign}
It follows that
\begin{equation}
    \label{eq:s-inequality}
    s^2 \leq \frac{1}{4} \left(1-\sqrt{1-10 n \epsilon ^2}\right) < 10\varepsilon^2 n,
\end{equation}
where the last inequality holds as long as $10\varepsilon^2 n < 1$, which we will assume from now on.

We want to know how far away $w_j^\prime$ is from any of the $\e^{\i\phi}w_i$ for any $\phi$ and any $i$ (because we do not care about ordering or overall phases). Thus, we want to know
\begin{salign}
    \max_j \min_{i,\phi} \lVert w_j^\prime - e^{i\phi}w_i\rVert^2
    &\leq 2\max_j \min_\phi (1-c_j' \cos\phi) \\
    &\leq 2\max_j (1-|c_j'|) \\
    &\leq 2\max_j (1-\sqrt{1-s_j^{\prime 2}}) \\
    &\leq 2 - 2\sqrt{1-s^2} \\
    (\text{\cref{eq:s-inequality}})
    &\leq 2 - 2 \sqrt{1 - 10\varepsilon^2 n} \\
    &< 20\varepsilon^2 n.
\end{salign}
We make the matrix \( V' \) from the \( w' \) vectors.
We see that there exists a phase matrix \( \Phi \) and permutation matrix \( P \) such that
\begin{equation}
    \norm{V' - W\Phi P}_{1,2}^2 < 20 \varepsilon^2 n,
\end{equation}
where the \( \norm{\cdot}_{1,2} \) norm (vector induced matrix norm) is the maximum column norm---that is, for any matrix $M$ and vector $\bm y$,
\begin{equation}
    \norm{M}_{1,2} = \max_{\bm x} \frac{\norm{M \bm x}_2}{\norm{\bm x}_1} = \max_j \sqrt{\sum_i M_{ij}^2},
    \quad 
    \norm{\bm y}_2 = \sqrt{\sum_i y_i^2},
    \quad 
    \norm{\bm y}_1 = \sum_i \abs*{y_i}.
\end{equation}
It follows that
for any matrix \( M \), 
\begin{equation}
    \norm{M} = \max_{\bm x} \frac{\norm{M \bm x}_2}{\norm{\bm x}_2}
    \leq \parentheses{\max_{\bm x}\frac{\norm{\bm x}_1}{\norm{\bm x}_2}} \parentheses{\max_{\bm x} \frac{\norm{M \bm x}_2}{\norm{\bm x}_1}}
    \leq \sqrt n \norm{M}_{1,2}.
\end{equation}
Thus, putting back the factor of \( 2/(b(b+1)) \) into \( \varepsilon \), there exists an unimportant \( \Phi P \) such that our estimate \( V' \) is
\begin{equation}
    \norm{V' - W\Phi P}_{\rm max} \leq \norm{V' - W\Phi P} < \frac{4\sqrt{5}\varepsilon n}{b(b+1)} \eqqcolon \delta
\end{equation}
where the max norm \( \norm{\cdot}_{\rm max} \) is the maximum absolute matrix entry and is always $\leq$ to the operator norm.

Therefore, $V^\prime$ will be
\(
V^\prime = W\Phi P + \delta B
\)
for some $\Phi$ and $P$ that we do not care about, and for some matrix $B$ with norm \( \leq 1 \), and therefore magnitude of entries $\leq 1$.
Note that \( C \coloneqq P^\dag \Phi^\dag W^\dag B \) has entries with magnitude \( \leq 1 \) as well.
We therefore have
\begin{equation}
    \abs{\bra{b^n} \calU_W^\dag \calU_{V'} \ket{b^n}}
    = \abs{\bra{b^n} \calU_P^\dag \calU_\Phi^\dag  \calU_W^\dag \calU_{V'} \ket{b^n}}
    = \abs{\bra{b^n} \calU_{\bbI + \delta C} \ket{b^n}}
    = \frac{1}{(b!)^n} \abs*{\perm(M)},
\end{equation}
where \( \perm \) denotes the matrix permanant and \( M \) is the \( bn\times bn \) matrix
\begin{equation}
    M = \begin{pmatrix}
        \bbI + \delta C & \dots & \bbI + \delta C \\
        \vdots          &       & \vdots          \\
        \bbI + \delta C & \dots & \bbI + \delta C
    \end{pmatrix}
    =
    \begin{pmatrix}
        \bbI   & \dots & \bbI   \\
        \vdots &       & \vdots \\
        \bbI   & \dots & \bbI
    \end{pmatrix} + \delta \begin{pmatrix}
        C      & \dots & C      \\
        \vdots &       & \vdots \\
        C      & \dots & C
    \end{pmatrix}
    .
\end{equation}
Using \cref{prop:large-permanent-perturbation} below, we can therefore bound
\begin{equation}
    \abs{\bra{b^n} \calU_W^\dag \calU_{V'} \ket{b^n}} \geq 1- \frac{nb\delta}{1-nb\delta},
\end{equation}
where \( \delta \) must satisfy \( \delta < (1/(bn)) \).
This completes the proof of \cref{thm:main-thm-const-fock}.

It therefore only remains to state and prove \cref{prop:large-permanent-perturbation}.

\begin{proposition}
    \label{prop:large-permanent-perturbation}
    Let \( E \) be a $bn\times bn$ matrix with entries between \( -1 \) and \( 1 \), and let \( I \) be the \( bn \times bn \) block matrix
    \begin{equation}
        I =
        \begin{pmatrix}
            \bbI_{n\times n} & \dots & \bbI_{n\times n} \\
            \vdots           &       & \vdots           \\
            \bbI_{n\times n} & \dots & \bbI_{n\times n}
        \end{pmatrix}.
    \end{equation}
    For any \( \varepsilon < 1/(bn) \), we have
    \begin{equation}
        \frac{1}{(b!)^n}\perm(I + \varepsilon E) \geq
        1 - \frac{\varepsilon bn}{1-\varepsilon bn}.
    \end{equation}
\end{proposition}
\begin{proof}
    Without loss of generality, assume that \( \varepsilon \geq 0 \).
    Let \( S_m \) denote the permutation group on \( m \) elements and \( \delta \) the Kronecker delta.
    We have that
    \begin{salign}
        \frac{1}{(b!)^n}\perm(I + \varepsilon E)
        &= \frac{1}{(b!)^n}\sum_{\pi \in S_{bn}} \prod_{i=1}^{bn} \parentheses{
            \delta_{i\equiv \pi(i) \bmod n}
            + \varepsilon E_{i,\pi(i)}
        } \\
        \begin{split}
            &= \frac{1}{(b!)^n} \sum_{\pi\in S_{bn}}\prod_{i=1}^{bn} \delta_{i\equiv \pi(i)\bmod n}
            + \frac{1}{(b!)^n}\varepsilon \sum_{i=1}^{bn} \sum_{\pi\in S_{bn}} E_{i,\pi(i)} \prod_{j\neq i} \delta_{j\equiv \pi(j) \bmod n} \\
            &\qquad + \frac{1}{(b!)^n}\varepsilon^2 \sum_{i < j=1}^{bn} \sum_{\pi\in S_{bn}} E_{i,\pi(i)}E_{j,\pi(j)} \prod_{\substack{k\neq i \\ k \neq j}} \delta_{k\equiv \pi(k)\bmod n} - \dots
        \end{split} \\
        \begin{split}
            &\geq 1
            - \frac{1}{(b!)^n}\varepsilon \sum_{i=1}^{bn} \sum_{\pi\in S_{bn}} \prod_{j\neq i} \delta_{j\equiv \pi(j) \bmod n} \\
            &\qquad - \frac{1}{(b!)^n}\varepsilon^2 \sum_{i < j=1}^{bn} \sum_{\pi\in S_{bn}} \prod_{\substack{k\neq i \\ k \neq j}} \delta_{k\equiv \pi(k)\bmod n} - \dots
        \end{split} \\
        &\geq 1 - \frac{(b!)^n}{(b!)^n} bn \varepsilon
        -
        \frac{(b!)^n}{(b!)^n} \binom{bn}{2}2! \varepsilon^2
        - \dots \\
        &= 1 - \sum_{k=1}^{bn} \frac{\varepsilon^k (bn)!}{(bn-k)!} \\
        &\geq 1 - \sum_{k=1}^{bn} (\varepsilon bn)^k
        \geq 1 - \sum_{k=1}^\infty (\varepsilon bn)^k
        = 1 - \frac{\varepsilon bn}{1-\varepsilon bn},
    \end{salign}
    completing the proof.
\end{proof}

\section{Description of the algorithm in Theorem S2}
\label{sec:ideal-description-arb-fock}

In this section, we prove \cref{thm:main-thm-arb-fock} in the error-free \( (\varepsilon_1 = \varepsilon_2 = 0) \) case by describing the full algorithm, which we summarize in \cref{alg:main-alg-arb-fock}.
In the next section, \cref{sec:noisy-description-arb-fock}, we prove the full theorem.

\begin{algorithm}[ht]
    \caption{Algorithm in \cref{thm:main-thm-arb-fock}}\label{alg:main-alg-arb-fock}
    \begin{algorithmic}[1]

        \Procedure{findVfock}{$\sigma^{(1)}, \sigma^{(2)}$}

        \State \( P_W \gets \sigma^{(1)} - \bbI \)
        \State Find \( U, \bm g \) such that \( P_W = U \operatorname{diag}(\bm g) U^\dag \) with \( g_i \leq g_{i+1} \)
        \State Round \( g_i \) to its nearest integer
        \State \( \tilde\sigma^{(2)} \gets (U^\dag \otimes U^\dag) \sigma^{(2)} (U\otimes U) \) \Comment{undo \( U \) to block diagonalize}

        \State Let \( 1=i_1<\dots<i_k=n \) be such that \( g_{i_j} = g_{i_j+1} = \dots = g_{i_{j+1}-1} \)
        \State \( X \gets \bbI_{n\times n} \)

        \For{\( j=1,\dots, k-1 \)}

        \State \( \ell \gets i_{j+1} - i_j \)
        \State Let \( \Gamma \) be the \( \ell^2 \times \ell^2 \) matrix \( \Gamma_{a,b;c,d} = \tilde\sigma^{(2)}_{i_j+a,i_j+b; i_j+c, i_j+d} \)
        \Comment{get the block moment matrices}

        \State \( V' =  \) \texttt{\sc findV}\( (\Gamma, g_{i_j}) \) ~~~ (from \cref{alg:main-alg-const-fock})
        \Comment{find the corresponding block unitary}

        \For{\( a,b=1,\dots, \ell \)}
        Set \( X_{i_j + a, i_j+b} = V'_{a,b} \)
        \EndFor

        \EndFor

        \State \( V \gets U X \) \Comment{redo \( U \)}

        \State \Return \( (V, \bm g) \)

        \EndProcedure

    \end{algorithmic}

\end{algorithm}

We therefore assume that we know both \( \sigma^{(1)} \) and \( \sigma^{(2)} \) perfectly for the state \( \calU_W\ket{\bm f} \) for unknown \( W \) and \( \bm f \).

Note that we can without loss of generality assume that \( f_i \leq f_{i+1} \) because any permutation can be absorbed into the unknown \( W \).
Before the application of \( W \), we have that
\begin{equation}
    \sigma^{(1)}_0 = \bbI + P, \qquad P = \operatorname{diag}(\bm f).
\end{equation}
Thus, with access to \( \sigma^{(1)} = W \sigma^{(1)}_0 W^\dag \), we can construct the matrix
\begin{equation}
    P_W \coloneqq  \sigma^{(1)} - \bbI = W P W^\dag.
\end{equation}
We diagonalize \( P_W \) so that \( P_W = U P U^\dag \) for some \( U \).
Thus, by diagonalizing, we have learned \( \bm f \).
However, \( U \) does not encode much about \( W \) because there is a lot of freedom in a \( U \) that diagonalizes \( P_W \).

However, note that
\begin{equation}
    \brackets{U^\dag W, P}
    = U^\dag W P - P U^\dag W
    = U^\dag W P - U^\dag P_W W
    = U^\dag W P - U^\dag W P
    = 0.
\end{equation}
Thus, \( U^\dag W \) consists of blocks of unitary matrices on the diagonal, where the blocks exactly correspond to the eigenspaces of \( P \).
Call the \( i^{\rm th} \) block \( \tilde W^{(i)} \) and let it be \( \ell_i\times \ell_i \).
We then have that
\( \tilde\sigma^{(2)} = (U^\dag \otimes U^\dag) \sigma^{(2)}(U\otimes U) = (U^\dag W \otimes U^\dag W) \sigma^{(2)}_0 (W^\dag U\otimes W^\dag U) \) will have the same block structure.
Note the \( i^{\rm th} \) smallest eigenvalue \( b_i \) of \( P \) will have multipilicty \( \ell_i \).
In other words, we will have
\begin{equation}
    b_1 = f_1 = \dots = f_{\ell_1}, \quad b_2 = f_{\ell_1+1} = \dots = f_{\ell_1+\ell_2}, \quad \text{etc.}
\end{equation}

Suppose we consider the \( i^{\rm th} \) block corresponding to eigenvalue \( b_i \) of \( P \), where this block is of size \( \ell_i \).
Then we can simply run \cref{alg:main-alg-const-fock} on the \( \ell_i^2\times \ell_i^2 \) matrix corresponding to the relevant block of \( \tilde\sigma^{(2)} \).
That is, we run \textsc{\tt findV} from \cref{alg:main-alg-const-fock} with the relevant block of \( \tilde\sigma^{(2)} \) and \( b_i \) as input.
From \cref{thm:main-thm-const-fock}, this will yield an \( \ell_i\times\ell_i \) unitary \( V^{(i)} \) such that \( \abs*{\bra{b_i^{\ell_i}}\calU_{\tilde W^{(i)}}^\dag \calU_{V^{(i)}} \ket{b_i^{\ell_i}}} = 1 \).
After doing this for all of the blocks, we define \( X = \oplus_i V^{(i)} \), and then \( V = U X \).
It then follows that
\begin{equation}
    \label{eq:repeated-applications}
    \abs{\bra{\bm f} \calU_W^\dag \calU_V \ket{\bm f}}
    = \abs{\bra{\bm f} \calU_W^\dag \calU_U \calU_X \ket{\bm f}}
    = \prod_{\text{block }i} \abs*{\bra{b_i^{\ell_i}} \calU_{\tilde W^{(i)}}^\dag \calU_{V^{(i)}} \ket{b_i^{\ell_i}} }
    = 1,
\end{equation}
as desired.

In summary, given access to \( \sigma^{(1)}, \sigma^{(2)} \) for a state \( \calU_W\ket{\bm f} \) for unknown \( W \) and \( \bm f \), we have found
a \( \bm g \) and a \( V \) such that \( \abs*{\bra{\bm f}\calU_W^\dag \calU_V \ket{\bm g}} = 1 \).
The procedure we just described is summarized explicitly in \cref{alg:main-alg-arb-fock}.

\section{Proof of Theorem S2}
\label{sec:noisy-description-arb-fock}

In this section, we prove \cref{thm:main-thm-arb-fock} by analyzing the effect that an error in our knowledge of the moments has on \cref{sec:ideal-description-arb-fock,alg:main-alg-arb-fock}.
We use the same notation as in the description of the algorithm in \cref{sec:ideal-description-arb-fock}, and thus all definitions carry over.

We assume that we have access to \( \sigma^{(1)\prime} = \sigma^{(1)} + \varepsilon_1 F \) and \( \sigma^{(2)\prime} = \sigma^{(2)} + \varepsilon_2 E \) where \( \norm{E}, \norm{F} \leq 1 \) .
It follows that in place of \( P_W \), we have \( P_W' = P_W + \varepsilon_1 F \).
As long as \( \varepsilon_1 < 1/2 \), after rounding we will find the correct \( \bm f \)
\cite[Thm.~VI.5.1,~VII.4.1]{bhatia1997matrix-analysis}.
Thus, we will have the correct block-diagonal structure.

Let \( \sigma^{(1)} = W \sigma^{(1)}_0 W^\dag = U \sigma^{(1)}_0 U^\dag \).
Note that \( \sigma^{(2)}_0 = (\sigma^{(1)}_0)^{\otimes 2} (\bbI + \text{SWAP}) - T \), where \( T = \sum_i f_i (f_i+1)\ket{i,i}\bra{i,i} \).
Let \( \sigma^{(1)\prime} = U' D' U^{\prime \dag} \).
From \cite[Below Thm.~VII.4.1]{bhatia1997matrix-analysis}, we have that
\begin{equation}
    \label{eq:eigval-bound-arb-fock}
    \norm*{\sigma^{(1)}_0 - D'} \leq \norm*{\sigma^{(1)} - \sigma^{(1)\prime}} \leq \varepsilon_1.
\end{equation}

Ultimately, we care about how close \( U^{\prime \dag}W \) is to a block-diagonal unitary. This is because, once it is block diagonal, we can appeal to \cref{thm:main-thm-const-fock} for the rest of the error bounds.
Thus, we will need the following lemma.

\begin{lemma}
    \label{lem:distance-to-block-unitary}
    Let \( D \) be a diagonal matrix with positive integer diagonal entries and let \( O\in \U(n) \).
    Then
    \begin{equation}
        \min_{\substack{V \in \U(n) \\ \text{s.t.~} [V, D] = 0}} \norm{O - V} \leq 2n \norm{[O, D]}.
    \end{equation}
\end{lemma}
\begin{proof}
    We write
    \begin{equation}
        D = \begin{pmatrix}
            d_1\bbI \\& d_2 \bbI \\&&\ddots
        \end{pmatrix},
        \quad
        O = \begin{pmatrix}
            O_{11} & O_{12} & \dots \\
            O_{21} & O_{22} & \dots \\
            \vdots & \vdots
        \end{pmatrix},
        \quad
        \tilde O = \begin{pmatrix}
            O_{11} & 0 \\ 0 & O_{22} \\&&\ddots
        \end{pmatrix},
        \quad
        V = \begin{pmatrix}
            V_1 \\ & V_2 \\&&\ddots
        \end{pmatrix},
    \end{equation}
    where all capital letters refer to matrices in the respective blocks.

    We have
    \begin{salign}
        \min_{\substack{V \in \U(n) \\ \text{s.t.~} [V, D] = 0}} \norm*{O - V}
        &\leq \norm*{O - \tilde O} + \min_{V_1, V_2,\dots} \norm*{\tilde O - V} \\
        %
        %
        &\leq \norm*{O - \tilde O} + \max_i \underbrace{\min_{U} \norm{O_{ii} - U}}_{\text{closest unitary to } O_{ii}} \\
        \text{(\cref{lem:distance-to-unitary})}
        &\leq
        \norm*{O - \tilde O} + \max_i \norm*{O_{ii}^\dag O_{ii}-\bbI}^{1/2} \\
        (\text{using }O^\dag O = \bbI)
        &\leq
        \norm*{O - \tilde O} + \norm{(O-\tilde O)^\dag(O-\tilde O)}^{1/2} \\
        &\leq
        2\norm*{O - \tilde O}.
    \end{salign}

    Thus, it only remains to lower bound \( \norm{[O, D]} \) by \( \norm*{O - \tilde O} \).
    Let \( \tilde D \) be the block matrix
    \begin{equation}
        \tilde D = \begin{pmatrix}
            0 & (d_2-d_1) \bbI & (d_3-d_1)\bbI &\dots \\
            -(d_2-d_1) \bbI & 0 & (d_3-d_2)\bbI & \dots \\
            -(d_3-d_1) \bbI & -(d_3-d_2)\bbI & 0 & \dots \\
            \vdots & \vdots & \vdots
        \end{pmatrix},
    \end{equation}
    and let \( \circ \) denote the Hadamard (entrywise) product.
    We have
    \begin{salign}
        \norm{[O, D]}
        &= \norm*{\tilde D \circ (O - \tilde O)} \\
        &\geq \norm*{\tilde D \circ (O - \tilde O)}_{\rm max} \\
        &\geq \norm*{O - \tilde O}_{\rm max} \\
        &\geq \frac{1}{n}\norm*{O - \tilde O},
    \end{salign}
    where we used the norm inequalities \( \norm{\cdot}_{\rm max} \leq \norm{\cdot} \leq n \norm{\cdot}_{\rm max} \)
\end{proof}

Above we used the following lemma.

\begin{lemma}
    \label{lem:distance-to-unitary}
    Let \( A \) be an $n\times n$ matrix. Then
    \begin{equation}
        \min_{O\in \U(n)}\norm{A - O} \leq \sqrt{\norm{A^\dag A - \bbI}}
        = \sqrt{\norm{A A^\dag - \bbI}}.
    \end{equation}
\end{lemma}
\begin{proof}
    Let \( A = U \Sigma V^\dag \) be the singular value decomposition of \( A \), so that \( U, V \in \U(n) \).
    Then
    \begin{equation}
        \min_{O\in \U(n)}\norm{A - O}
        \leq \norm{A - UV^\dag}
        = \norm{\Sigma - \bbI}
        \leq \sqrt{\norm{\Sigma^2 - \bbI}}
        = \sqrt{\norm{A^\dag A - \bbI}}
        = \sqrt{\norm{A A^\dag - \bbI}}.
    \end{equation}
\end{proof}

To use \cref{lem:distance-to-block-unitary}, we need to bound the commutator, or equivalently, determine how close \( U^{\prime \dag}W \sigma^{(1)}_0 (U^{\prime \dag}W)^\dag \) is to \( \sigma^{(1)}_0 \):
\begin{salign}
    \norm{U^{\prime \dag} \sigma^{(1)} U^{\prime} - \sigma^{(1)}_0}
    &= \norm{U^{\prime \dag} \sigma^{(1)\prime} U^{\prime} - \sigma^{(1)}_0 - \varepsilon_1 U^{\prime \dag} F U^{\prime}} \\
    &= \norm{D' - \sigma^{(1)}_0 - \varepsilon_1 U^{\prime \dag} F U^{\prime}} \\
    &\leq \norm{D' - \sigma^{(1)}_0} + \varepsilon_1 \norm{F} \\
    (\text{\cref{eq:eigval-bound-arb-fock}})
    &\leq 2 \varepsilon_1.
\end{salign}
By \cref{lem:distance-to-block-unitary}, this also bounds the distance to the nearest block-diagonal unitary matrix, with an additional factor of \( 2n \).
Thus, we let \( O \) be the nearest block-diagonal unitary matrix, and we have\footnote{The $n$ dependence in \cref{eq:block-distance-arb-fock} can be removed. Indeed in the proof of \cref{thm:main-thm-gaussian}, which is more general than the proof here and so also applies to the proof here, we have no $n$ dependence in the analogous part of the proof, \cref{eq:distance-to-passive-gaussian}. However, we leave this here as is because it is direct and allows us to keep track of all constant factors.}
\begin{equation}
    \label{eq:block-distance-arb-fock}
    \norm{U^{\prime \dag} W - O} \leq 4n\varepsilon_1.
\end{equation}
Then,
\begin{salign}
    \norm{\tilde\sigma^{(2)\prime} - O^{\otimes 2} \sigma^{(2)}_0 O^{\dag \otimes 2}}
    &= \norm{U^{\prime \dag \otimes 2}\sigma^{(2)\prime} U^{\prime \otimes 2} - O^{\otimes 2} \sigma^{(2)}_0 O^{\dag \otimes 2}} \\
    &= \norm{U^{\prime \dag \otimes 2}\sigma^{(2)} U^{\prime \otimes 2} - O^{\otimes 2} \sigma^{(2)}_0 O^{\dag \otimes 2} + \varepsilon_2 U^{\prime \dag \otimes 2} E U^{\prime \otimes 2}} \\
    &\leq \norm{U^{\prime \dag \otimes 2}\sigma^{(2)} U^{\prime \otimes 2} - O^{\otimes 2} \sigma^{(2)}_0 O^{\dag \otimes 2}} + \varepsilon_2 \norm{E} \\
    &\leq \norm{(U^{\prime \dag} W)^{\otimes 2}\sigma^{(2)}_0 (W^\dag U^{\prime})^{\otimes 2} - O^{\otimes 2} \sigma^{(2)}_0 O^{\dag \otimes 2}} + \varepsilon_2 \\
    (\text{\cref{eq:telescoping-sum}})
    &\leq
    4\norm{U^{\prime \dag} W - O} \norm*{\sigma^{(2)}_0}
    + \varepsilon_2 \\
    (\text{\cref{eq:block-distance-arb-fock}})
    &\leq
    16n\varepsilon_1 \norm*{\sigma^{(2)}_0}
    + \varepsilon_2 \\
    &\leq
    16n\varepsilon_1 \parentheses{2\norm*{\sigma_0^{(1)}}^2 + \norm{T}}
    + \varepsilon_2 \\
    &=
    16n\varepsilon_1 \parentheses{2(1+f_{\rm max})^2 + f_{\rm max}(f_{\rm max}+1)}
    + \varepsilon_2 \\
    &= 16 n\varepsilon_1 (3 f_{\rm max}^2 + 5 f_{\rm max} + 2) + \varepsilon_2  \eqqcolon \delta,
\end{salign}
where we used the telescoping sum
\begin{equation}
    \label{eq:telescoping-sum}
    \begin{split}
        (A\otimes A) \Lambda (A\otimes A)^T - (O\otimes O) \Lambda (O\otimes O)^T
        &= ((A-O)\otimes A)\Lambda (A\otimes A)^T + (O \otimes (A-O))\Lambda (A\otimes A)^T \\
        &~+ (O \otimes O)\Lambda ((A-O)\otimes A)^T + (O \otimes O)\Lambda (O\otimes (A-O))^T.
    \end{split}
\end{equation}

Thus, we are effectively sending the blocks of \( O^{\otimes 2} \sigma^{(2)}_0 O^{\dag \otimes 2} \) with some error into \cref{alg:main-alg-const-fock}.
Because the norm of a submatrix is upper bounded by that of the matrix, each of the blocks that we send into \cref{alg:main-alg-const-fock} will be the relevant block of \( O^{\otimes 2} \sigma^{(2)}_0 O^{\dag \otimes 2} \) plus some error \( G \) with \( \norm{G} \leq \delta = 16 n\varepsilon_1 (3 f_{\rm max}^2 + 5 f_{\rm max} + 2) + \varepsilon_2 \).

Using \cref{thm:main-thm-const-fock},
this will give us an estimate \( \tilde O \) of \( O \) that satisfies
\begin{equation}
    \label{eq:Otilde-error-arb-fock}
    \norm*{\tilde O - O\Phi P} \leq 2\sqrt5 n \delta,
\end{equation}
where \( \Phi P \) is an unimportant block-diagonal phase and permutation.
\( V = U' \tilde O \) is therefore our estimate of \( W \Phi P \), and it satisfies
\begin{salign}
    \norm*{V - W \Phi P}
    &= \norm*{\tilde O - U^{\prime \dag}W \Phi P} \\
    &\leq \norm*{\tilde O - O \Phi P} + \norm*{O \Phi P - U^{\prime \dag}W \Phi P} \\
    (\text{\cref{eq:Otilde-error-arb-fock}})
    &\leq 2\sqrt5 n \delta + \norm*{O - U^{\prime \dag}W} \\
    (\text{\cref{eq:block-distance-arb-fock}})
    &\leq 2\sqrt5 n \delta + 4n\varepsilon_1,
\end{salign}
completing the first part of \cref{thm:main-thm-arb-fock}.

An easy generalization of the proof of \cref{prop:large-permanent-perturbation} yields
%
\begin{equation}
    \abs{\bra{\bm f} \calU_W^\dag \calU_V \ket{\bm f}} \geq 1 - \frac{(2\sqrt5 n \delta + 4n\varepsilon_1) f_{\rm max} n}{1 - (2\sqrt5 n \delta + 4n\varepsilon_1) f_{\rm max} n}
\end{equation}
as long as \( (2\sqrt5 n \delta + 4n\varepsilon_1) f_{\rm max} n < 1 \),
completing the proof of \cref{thm:main-thm-arb-fock}.

\section{Description of the algorithm in Theorem S3}
\label{sec:ideal-description-gaussian}

In this section, we prove \cref{thm:main-thm-gaussian} in the error-free \( (\varepsilon_1 = \varepsilon_2 = 0) \) case by describing the full algorithm, which we summarize in \cref{alg:main-alg-gaussian}.
In the next section, \cref{sec:noisy-description-gaussian}, we prove the full theorem.

\begin{algorithm}[ht]
    \caption{Algorithm in \cref{thm:main-thm-gaussian}}\label{alg:main-alg-gaussian}
    \begin{algorithmic}[1]

        \setstretch{1.35}

        \Procedure{findQ}{$\Lambda^{(1)}, \Lambda^{(2)}$}

        \State \( (\bm\nu, R) \gets  \) Williamson decomposition of \( \Re\Lambda^{(1)} \) \Comment{\( \bm\nu \) are the symplectic eigenvalues}
        \State \( \bm g \gets \bm\nu - 1/2 \)

        \State \( \tilde \Lambda^{(2)} \gets (R^{-1}\otimes R^{-1}) \Lambda^{(2)} (R^{-1}\otimes R^{-1})^T \) \Comment{undo active part of the Gaussian unitary}


        \State \( \tilde\sigma^{(2)}_{ij;kl} \gets \frac{1}{4}\sum_{a,b,c,d=0}^1 \i^{a+b}(-\i)^{c+d} \tilde\Lambda^{(2)}_{i+na, j+nb; k+nc; l + nd} \) \Comment{convert from \( x, p \)'s to \( a,a^\dag \)'s}

        \State Run lines 6-13 of \cref{alg:main-alg-arb-fock} on $\tilde\sigma^{(2)}$ to find $V$ \Comment{runs \cref{alg:main-alg-const-fock} on each block}

        \State \( O \gets \rho(V) = \begin{pmatrix}
            \Re V & -\Im V \\ \Im V & \Re V
        \end{pmatrix} \) \Comment{\cref{eq:passive-isomorphism}}
        \State \( Q \gets R O \) \Comment{reapply the active part}
        \State \Return \( (Q, \bm g) \)

        \EndProcedure

    \end{algorithmic}

\end{algorithm}

We therefore assume that we know \( \Lambda^{(1)} \) and \( \Lambda^{(2)} \) perfectly for the state \( \calU_S \ket{\bm f} \).
We consider the initial Fock state \( \ket{\bm f} \).
Define the matrix \( P_{\bm f} = \operatorname{diag}\cbargs{f_1,\dots, f_n} \).
Using
\begin{equation}
    \label{eq:a-adag-x-p}
    a = \frac{x+\i p}{\sqrt 2}, \quad a^\dag = \frac{x - \i p}{\sqrt 2}, \quad x = \frac{a+a^\dag}{\sqrt 2}, \quad p = \frac{\i(a^\dag - a)}{\sqrt 2},
\end{equation}
we have
\begin{salign}[eq:Lambda-1-1]
    \Lambda^{(1)}_0 &=  \sum_{i=1}^{n} \parentheses{ \frac{1}{2} + f_{i}} \parentheses{\ket{i} \bra{i} + \ket{n+i} \bra{n+i} + \frac{\i}{2} \ket{i}\bra{n+i} - \frac{\i}{2} \ket{n+i}\bra{i} } \\
    &= \begin{pmatrix}
        \frac{1}{2}\bbI + P_{\bm f} & 0                           \\
        0                           & \frac{1}{2}\bbI + P_{\bm f}
    \end{pmatrix} + \frac{\i}{2}\sum_{i=1}^n (1+f_i) \parentheses{ \ket{i}\bra{n+i} - \ket{n+i}\bra{i} }.
\end{salign}
As before, we can, without loss of generality, assume that \( f_i \leq f_{i+1} \).

Notice that \( \frac{1}{2}+\bm f \) are the symplectic eigenvalues of \( \Re \Lambda_0^{(1)} \) (\emph{i.e.}~the positive eigenvalues of \( \i \Omega \Re \Lambda^{(1)} \), where \( \Omega \) is the symplectic form) \cite{serafini2017quantum-continu}.
Symplectic eigenvalues do not change under the application of a symplectic matrix.
Thus,
given access to \( \Lambda^{(1)} \), we can determine \( \bm f \) by computing the symplectic eigenvalues of \( \Re \Lambda^{(1)} \).
In particular, we perform the Williamson decomposition \cite{serafini2017quantum-continu} on \( \Lambda^{(1)} \) to get the symplectic eigenvalues \( \bm \nu \) and a symplectic diagonalizing matrix \( R \).
It follows that
\begin{equation}
    \Re \Lambda^{(1)} = S (\Re \Lambda^{(1)}_0 ) S^T = R (\Re \Lambda^{(1)}_0 ) R^T.
\end{equation}
Because \( S \) and \( R \) both diagonalize the same covariance matrix, \( R^{-1}S \in K(n) \) and so corresponds to a passive Gaussian unitary \cite[Prop.~8.12]{gosson2006symplectic-geom}.
In fact, $R^{-1}S$ corresponds to a block diagonal passive Gaussian unitary, where the block diagonal is with respect to the degeneracy of the symplectic eigenvalues \cite{Babu_Mishra_2024}.
It follows that with
\begin{equation}
    \tilde \Lambda^{(1)} = R^{-1} \Lambda^{(1)} (R^{-1})^T,
    \qquad
    \tilde\Lambda^{(2)} = (R^{-1})^{\otimes 2} \Lambda^{(2)} (R^{-1 \otimes 2})^T,
\end{equation}
\( \tilde\Lambda^{(1)} \) and \( \tilde\Lambda^{(2)} \) are the moment matrices for the initial Fock state \( \bm f \) acted on by an unknown block diagonal passive Gaussian unitary specified by a \( W \in \U(n) \) with \( \rho(W) = R^{-1}S \), where \( \rho \) is the isomorphism in \cref{eq:passive-isomorphism}.
Thus, each block falls into the setting of \cref{thm:main-thm-const-fock}, and we can therefore use \cref{alg:main-alg-const-fock} to find \( V \) (which acts equivalently to \( W \)).
We define the symplectic matrix \( Q = R \rho(V) \).
Then \( \calU_Q \) acts equivalently to \( \calU_S \) on the initial state.

Note that, in order to use \cref{alg:main-alg-const-fock} with \( \tilde\Lambda^{(1)} \) and \( \tilde\Lambda^{(2)} \), we need to convert to the \( \sigma \)-type moment matrices (in fact, we only need to use the $\sigma^{(2)}$ moments for \cref{alg:main-alg-const-fock}, but we show how to convert both for completeness).
Using \cref{eq:a-adag-x-p}, this is simply
\begin{salign}[eq:convert-to-sigma]
    \sigma^{(1)}_{ij} &= \frac{1}{2}\parentheses{\tilde \Lambda^{(1)}_{ij} + \tilde \Lambda^{(1)}_{n+i,n+j} + \i \tilde \Lambda^{(1)}_{n+i, j} - \i \tilde \Lambda^{(1)}_{i,n+j} } \\
    \sigma^{(2)}_{ij;kl} &= \frac{1}{4}\sum_{a,b,c,d=0}^1 \i^{a+b}(-\i)^{c+d} \tilde\Lambda^{(2)}_{i+na, j+nb; k+nc; l + nd} .
\end{salign}
Thus, we arrive at \cref{alg:main-alg-gaussian}.

\section{Proof of Theorem S3}
\label{sec:noisy-description-gaussian}

In this section, we prove \cref{thm:main-thm-gaussian} by analyzing the effect that an error in our knowledge of the moments has on \cref{sec:ideal-description-gaussian,alg:main-alg-gaussian}.
We use the same notation as in the description of the algorithm in \cref{sec:ideal-description-gaussian}, and thus all definitions carry over.

\subsection{Preliminary lemmas}

Before proving \cref{thm:main-thm-gaussian}, we first state and prove a number of other useful results.
On a first readthrough, we recommend reading the proof in \cref{sec:proof-of-gaussian} prior to reading this section.


\begin{lemma}
    \label{lem:norm-of-symplectic-inverse}
    Let \( S \in \Sp(2n) \). Then \( \norm{S} = \norm*{S^{-1}} \).
\end{lemma}
\begin{proof}
    The Euler (\emph{a.k.a.}~Block-Messiah) decomposition gives \( S = O \diag(A, A^{-1}) V \) for \( O, V \in \O(2n) \) \cite{serafini2017quantum-continu}.
    Therefore, \( \norm*{S^{-1}} = \norm*{V^T \diag(A^{-1}, A) O^T} = \max(\norm*{A}, \norm*{A^{-1}}) = \norm{S} \).
\end{proof}

The next theorem is taken directly from Ref.~\cite{idel2017perturbation-bo}. It proves stability of symplectic eigenvalues under perturbations.

\begin{theorem}[Theorem 3.1, Eq.~(5) of Ref.~\cite{idel2017perturbation-bo}]
    Let \( M, M' \in \bbR^{2n\times 2n} \) be positive-definite matrices and let \( D,D' \in \bbR^{n\times n} \) be the nonnegative diagonal matrices corresponding to the symplectic eigenvalues of \( M, M' \) (\emph{i.e.}~\( S^T M S = \diag(D, D) \) for some \( S \in \Sp(2n, \bbR) \), and similarly for \( M' \)).
    Then
    \begin{equation}
        \nnorm{D - D'} \leq \sqrt{\kappa(S)\kappa(M')} \nnorm*{M - M'}
        ,
    \end{equation}
    where \( \kappa(X) \) denotes the condition number of \( X \), and \( \nnorm{\cdot} \) denotes any unitarily invariant norm.
\end{theorem}

We now derive the corresponding bound in the setting of \cref{thm:main-thm-gaussian}.

\begin{corollary}
    \label{cor:symp-eig-errors}
    Suppose that a covariance matrix \( M \) is known up to some error matrix \( F \),
    where we assume that \( M \) corresponds to a state beginning in a Fock state \( \bm f \) and acted upon by a Gaussian unitary.
    The symplectic eigenvalues \( D \) of \( M \) and the symplectic eigenvalues \( D' \) of \( M + F \) are related by
    \begin{equation}
        \norm{D - D'} \leq \e^{s} \sqrt{2(1+2\max_i f) \e^{2s} \norm{F}^2 + 4\norm{F}^3},
    \end{equation}
    as long as \( \norm{F} \leq 1/4 \).
    Here \( \norm{\cdot} \) denotes the operator norm, and \( s \) denotes the maximum squeezing in the Euler decomposition \cite{serafini2017quantum-continu} of the symplectic diagonalizing matrix $S$ of \( M \) (that is, $\e^s$ is the maximum singular value of $S$).
\end{corollary}
\begin{proof}
    Corollary of the above theorem along with \( \kappa(AB)\leq \kappa(A)\kappa(B) \) for square matrices \cite{551135}.

    We simply need to understand the condition numbers of \( S \), where \( M = S \diag(D,D)S^T \), and \( M' = M + F \).
    Note that \( M' = R \diag(D', D')R^T \) for some symplectic matrix \( R \).
    By the Euler decomposition of a symplectic matrix, we have that
    \begin{equation}
        S = O A U,
        \qquad
        M' = O' A' U' \diag(D', D') (U')^T A' (O')^T,
    \end{equation}
    where \( A = \diag(e^{\bm s}, \e^{-\bm s}) \), \( A' = \diag(e^{\bm s'}, \e^{-\bm s'}) \), \( \bm s = (s_1,\dots,s_n) \) denotes the squeezing operation that \( S \) contains,
    and \( O,U, O', U' \) are orthogonal symplectic matrices corresponding to passive Gaussian unitaries.

    Because the condition number of an orthogonal matrix is \( 1 \), we have that
    \( \kappa(S) = \e^{2s} \).
    Furthermore, \( \kappa(M) \leq \e^{2s} d \), where \( s = \max_i \abs{s_i} \), \( d = (\max_i D_{ii}) / (\min_i D_{ii}) \leq \frac{1/2+\max_i f_i}{1/2} = 1 + 2\max_i f_i \).

    We also have that
    \begin{salign}
        \kappa(M')
        &= \kappa(M + F) \\
        &\leq \frac{\text{max eigval of } M + \norm{F}}{\text{min eigval of } M - \norm{F}} \\
        &\leq \frac{2\norm{S \diag(D, D) S^T} + 2\norm{F}}{1-2\norm{F}} \\
        &\leq \frac{(1+2\max_i f) \e^{2s} + 2\norm{F}}{1-2\norm{F}}.
    \end{salign}

    Thus, we have that
    \begin{salign}
        \norm{D - D'}
        &\leq \sqrt{\kappa(S)\kappa(M')} \norm{F} \\
        &\leq \e^{s} \sqrt{\frac{(1+2\max_i f) \e^{2s} + 2\norm{F}}{1-2\norm{F}}} \norm{F} \\
        &= \e^{s} \sqrt{\frac{(1+2\max_i f) \e^{2s} \norm{F}^2 + 2\norm{F}^3}{1-2\norm{F}}} \\
        \parentheses{\text{assume } \norm{F} \leq \frac{1}{4}}
        &\leq
        \e^{s} \sqrt{2(1+2\max_i f) \e^{2s} \norm{F}^2 + 4\norm{F}^3}.
    \end{salign}
\end{proof}

Finally, given a symplectic matrix, we will need to determine how close it is to one corresponding to a block diagonal passive Gaussian unitary.
Intuitively, above we used stability results on the symplectic eigenvalues, whereas now we will need to use stability results on the diagonalizing symplectic matrix.
Above, we used the error bounds coming from Ref.~\cite{idel2017perturbation-bo} for the symplectic eigenvalues.
In a previous version of this paper (arXiv version 1), we also used their bounds for the symplectic diagonalizing matrix. However, for the symplectic diagonalizing matrix, we can do better with Ref.~\cite{Babu_Mishra_2024}.

\begin{theorem}[Thm.~6 of \cite{Babu_Mishra_2024}, restated]
\label{thm:babu_mishra}
Let $A$ be a $2n\times 2n$ positive definite matrix and $H$ a symmetric perturbation so that $A+H$ is positive definite.
Let $S$ symplectically diagonalize $A$ and $S'$ symplectically diagonalize $A+H$.
Then there exists a block diagonal orthogonal symplectic matrix $O$ such that $S' = SO + \bigO{\norm{H}}$.
Here, block diagonal refers to the blocks of equal symplectic eigenvalues.
\end{theorem}

This theorem tells us that, in the error free case of our learning algorithm when we know $\Lambda^{(2)}$ perfectly, when we symplectically diagonalize it to find $R$, our $R$ satisfies $R^{-1}S = O$ for some block diagonal passive Gaussian unitary $O$.
We can therefore send each of the individual blocks of $R^{-1\otimes t}\Lambda^{(t)} R^{-T \otimes t}$ into \cref{alg:main-alg-const-fock}.
If $O$ were not block diagonal but instead just a passive Gaussian unitary, then we would instead have to send $R^{-1\otimes t}\Lambda^{(t)} R^{-T \otimes t}$ into \cref{alg:main-alg-arb-fock} (which is what we did in the arXiv version V1 of this paper).
Fortunately, the fact that $O$ is already block diagonal immediately improves the error bounds of the algorithm.
With errors, so that we have $R'$ instead of $R$, we will therefore need to understand how far away $R^{\prime-1}S$ is from such an $O$.
This is \cref{cor:babu_mishra}.

\begin{corollary}
    \label{cor:babu_mishra}
    Suppose that $M$ is the covariance matrix for the state $\calU_S \ket{\bm f}$ and $M' = M + \varepsilon F$ is a covariance matrix with $\norm{F} \leq 1$. Let $R'$ symplectically diagonalize $M'$.
    Define $\e^s$ as the max singular value of $S$ and $f_{\rm max} = \max_i f_i$.
    Then
    \begin{equation}
        \min_{\substack{O \in \Sp(2n)\cap \O(2n) \\ \text{s.t.~$O$ is block diagonal}}} \norm{R^{\prime-1}S - O} \leq \bigO{\e^s \varepsilon}.
    \end{equation}
\end{corollary}
\begin{proof}
    From \cref{thm:babu_mishra}, there exists a block diagonal passive Gaussian unitary $O$ such that $R' = SO + P$ with $\norm P \leq \bigO{\varepsilon}$.
    It follows that the quantity of interest is upper bounded by
    \begin{salign}
        \norm{R^{\prime-1}S - O}
        &\leq \norm*{R^{\prime-1}}\norm{S - R' O} \\
        &\leq \norm*{R^{\prime-1}} \varepsilon \\
        (\text{\cref{lem:norm-of-Rprime}})
        &\leq \e^s \varepsilon + \bigO{\varepsilon^2}.
        %
    \end{salign}
\end{proof}

Above, we used the following.

\begin{lemma}
    \label{lem:norm-of-Rprime}
    Suppose that \( M \) is a covariance matrix and \( M' = M + \varepsilon F \) is a covariance matrix with \( \norm{F} \leq 1 \). Let  \( R' \) symplectically diagonalize \( M' \) and $S$ symplectically diagonalize $M$.
    Define \( \e^{s} \) as the max singular value of \( S \) and \( f_{\rm max} = \max_i f_i \).
    Then
    \begin{equation}
        \norm{R'} = \norm*{R^{\prime -1}} \leq \e^s + \bigO{\varepsilon}.
    \end{equation}
\end{lemma}
\begin{proof}
    We prove the proposition for \( R^{\prime} \), and the full proposition follows from \cref{lem:norm-of-symplectic-inverse}.
    From \cref{thm:babu_mishra}, there exists an orthogonal $O$ such that $R' = SO + \bigO{\varepsilon}$. 
    Therefore, $\norm{R'} \leq \norm{S}\norm{O} + \bigO{\varepsilon} = \e^s + \bigO{\varepsilon}$.
\end{proof}

\subsection{Proof of Theorem S3}
\label{sec:proof-of-gaussian}

We use the same notation as in the description of the algorithm in \cref{sec:ideal-description-gaussian}, and thus all definitions carry over.
Let \( M = \Re \Lambda^{(1)} \) and \( M' = \Re \Lambda^{(1)\prime} = M + \varepsilon_1 F \).
Let \( M = R \nu R^T = S \nu S^T \),  \( M' = R' \nu' R^{\prime T} \) be a symplectic diagonalization of \( M, M' \), where \( \nu = \diag(D, D) \), \( \nu' = \diag(D', D') \).
From \cref{cor:symp-eig-errors}, as long as $\varepsilon_2 \leq \bigO{\frac1{\e^{2s}f_{\rm max}}}$, which we will assume from now on, we can round $\nu'$ to the nearest integer to get the true $\nu$.
Thus, we will learn the correct initial Fock state.

As described in \cref{sec:ideal-description-gaussian},
in the ideal case when \( \varepsilon_1 = 0 \) so that \( R = R' \), \( R^{\prime -1} S \) is an orthogonal matrix that is block diagonal with respect to the symplectic eigenvalue degeneracy.
Therefore, \( \tilde\Lambda^{(i)} \) are the correlation matrices associated to a passive Gaussian block unitary applied to a Fock state.
In this case, if we know \( \tilde\Lambda^{(i)} \) up to an error, then we know how the error propagates due to our analysis in \cref{thm:main-thm-const-fock}.
Therefore, for the analysis in this section when \( \varepsilon_i > 0 \), we need to first understand how close \( R^{\prime -1}S \) is to an orthogonal matrix.
For this, we use \cref{cor:babu_mishra}, which tells us that there exists a passive block diagonal Gaussian unitary specified by $O \in \Sp(2n)\cap O(2n)$ satisfying
\begin{equation}
    \label{eq:distance-to-passive-gaussian}
    \norm*{R^{\prime -1}S - O} \leq \delta = \bigO{\e^s \varepsilon_1}.
\end{equation}
where \( O \) is some passive Gaussian unitary.

We suppose that \( \Lambda^{(2)\prime} = \Lambda^{(2)} + \varepsilon_2 E = S^{\otimes 2}\Lambda^{(2)}_0 S^{T \otimes 2} + \varepsilon_2 E \), with \( \norm{E} \leq 1 \).
Again, following analogously to the proof of \cref{thm:main-thm-arb-fock}, we wish to bound
\begin{salign}
    \norm{\tilde\Lambda^{(2)\prime} - O^{\otimes 2}\Lambda^{(2)}_0 O^{T \otimes 2}}
    &= \norm{(R^{\prime-1}S)^{\otimes 2}\Lambda^{(2)}_0 (R^{\prime-1} S)^{T \otimes 2} + \varepsilon_2 (R^{\prime-1})^{\otimes 2}E (R^{\prime-1})^{T\otimes 2} - O^{\otimes 2}\Lambda^{(2)}_0 O^{T \otimes 2}} \\
    &\leq \norm{(R^{\prime-1}S)^{\otimes 2}\Lambda^{(2)}_0 (R^{\prime-1} S)^{T \otimes 2} - O^{\otimes 2}\Lambda^{(2)}_0 O^{T \otimes 2}} + \varepsilon_2 \norm{R^{\prime-1}}^4 \norm{E} \\
    (\text{\cref{lem:norm-of-Rprime}})
    &\leq \norm{(R^{\prime-1}S)^{\otimes 2}\Lambda^{(2)}_0 (R^{\prime-1} S)^{T \otimes 2} - O^{\otimes 2}\Lambda^{(2)}_0 O^{T \otimes 2}} + \bigO{\varepsilon_2 \e^{4s}} \\
    \begin{split}
        (\text{\cref{eq:telescoping-sum}})
        &\leq \norm{R^{\prime -1}S - O} \norm*{\Lambda_0^{(2)}} \parentheses{\norm{R^{\prime -1}S}^3+\norm{R^{\prime -1}S}^2+\norm{R^{\prime -1}S}+1} \\
        &\qquad + \bigO{\varepsilon_2 \e^{4s}}
    \end{split}\\
    (\text{\cref{eq:distance-to-passive-gaussian}})
    &\leq \delta \norm*{\Lambda_0^{(2)}} \parentheses{(1+\delta)^3+(1+\delta)^2+(1+\delta)+1} + \bigO{\varepsilon_2 \e^{4s}} \\
    &\leq \bigO{\delta f_{\rm max}^2 + \varepsilon_2 \e^{4s}} \\
    &= \bigO{\varepsilon_1 \e^s  f_{\rm max}^2 + \varepsilon_2 \e^{4s}} \eqqcolon \eta.
\end{salign}

Now we create the $\sigma^{(2)}$ matrix from $\tilde\Lambda^{(2)\prime}$ via \cref{eq:convert-to-sigma}.
Because this is just a unitary transformation and then taking a sub-block, we have
\begin{equation}
    \norm*{\sigma^{(2)} - U^{\otimes 2}\sigma^{(2)}_0 U^{\dag \otimes 2}} \leq \eta
\end{equation}
for some block diagonal unitary $U$.

We now proceed exactly as in the end of \cref{sec:noisy-description-arb-fock}.
We are sending the blocks of $U^{\otimes 2}\sigma^{(2)}_0 U^{\dag \otimes 2}$ plus some error into \cref{alg:main-alg-const-fock}.
As already examined there, this yields an estimate $\tilde U$ of $U$ that satisfies
\begin{equation}
    \norm*{\tilde U - U\Phi P} \leq \delta_1 = \bigO{n \eta},
\end{equation}
and thus we find a \( \tilde O \) such that
\begin{equation}
    \norm*{\tilde O - O \rho(\Phi P)} \leq \delta_1,
\end{equation}
where \( \Phi \) and \( P \) are the unimportant phase and permutation unitary matrices and $\rho$ is as in \cref{eq:passive-isomorphism}.
Our final estimate of \( S \) up to phases and permutations is \( Q = R' \tilde O \). We have that
\begin{salign}
    \norm{Q - S \Phi P}
    &= \norm*{R' \tilde O - S \Phi P} \\
    &\leq \norm*{R' (\tilde O - O \Phi P)} + \norm{R' O \Phi P - S \Phi P} \\
    &\leq \norm{R'} \delta_1 + \norm{R' O - S} \\
    &\leq \norm{R'} \parentheses{\delta_1 + \norm{O - R^{\prime -1}S}} \\
    (\text{\cref{eq:distance-to-passive-gaussian}})
    &\leq \norm{R'} \parentheses{\delta_1 + \delta} \\
    (\text{\cref{lem:norm-of-Rprime}})
    &\leq \bigO{\e^s \parentheses{\delta_1 + \delta}} \\
    %
    &\leq \bigO{
    \varepsilon_1 n \e^{2s}  f_{\rm max}^2 + \varepsilon_2 n \e^{5s}
    } \eqqcolon \delta_2,
\end{salign}
completing the first part of the proof of \cref{thm:main-thm-gaussian}.

By extension,
\begin{equation}
    \norm{P^T \Phi^T S^{-1}Q - \bbI} \leq \norm{S} \norm{Q-S\Phi P}
    \leq \delta_3 \coloneqq \delta_2 \e^s.
\end{equation}
We want to bound
\( \abs{\bra{\bm f}\calU_S^\dag \calU_Q \ket{\bm f}} = \abs{\bra{\bm f}\calU_P^\dag \calU_\Phi^\dag \calU_S^\dag \calU_Q \ket{\bm f}} \).
Therefore, the last task is to prove the following proposition.

\begin{proposition}
    Given \( X \) with \( \norm{X - \bbI} \leq \gamma \), it follows that
    \begin{equation}
        \abs{\bra{\bm f} \calU_X \ket{\bm f}} \geq 1 - \bigO{n \gamma f_{\rm max}}.
    \end{equation}
\end{proposition}
\begin{proof}
    Without changing the results of the analysis, we can switch basis such that instead of \( r = (x, p) \) we can let \( r = (a^\dag, a) \).
    As usual with a Gaussian unitary specified by \( X \), it acts as \( r_I \mapsto X_{IJ}r_J \), where little indices will go from \( 1,\dots,n \) and big indices will go from \( 1,\dots, 2n \).
    Let \( \gamma M = X - \bbI \).
    Therefore, we have that
    \begin{equation}
        a_i^\dag \mapsto X_{i J} r_J = a_i^\dag + \gamma(M_{iJ}r_J).
    \end{equation}
    Therefore,
    \begin{salign}
        \abs*{\bra{\bm f} \calU_X \ket{\bm f}}
        &= \abs*{\bra{\bm f} \prod_i \parentheses{\frac{a_i^\dag + \gamma M_{iJ}r_J}{\sqrt{ f_i!}}}^{f_i} \ket0 } \\
        &\geq 1 - \bigO{n \gamma f_{\rm max}}.
    \end{salign}
\end{proof}

Thus, we have that
\begin{equation}
    \abs{\bra{\bm f}\calU_S^\dag \calU_Q \ket{\bm f}} \geq 1 - n f_{\rm max} \delta_3
    = 1 -
    \bigO{
    \varepsilon_1 n^2 \e^{3s}  f_{\rm max}^3 + \varepsilon_2 n^2 \e^{6s} f_{\rm max}
    },
\end{equation}
completing the proof of \cref{thm:main-thm-gaussian}.



\section{Measuring correlation matrices and the full end-to-end sample complexity}
\label{sec:measurement}

In \cref{thm:main-thm-const-fock,thm:main-thm-arb-fock,thm:main-thm-gaussian}, we derive a learning algorithm assuming that one is able to measure the matrix elements of \( \sigma^{(t)} \) and \( \Lambda^{(t)} \) to inverse polynomial norm precision.
In general, this can be done by using only Gaussian measurements.
Specifically, we sample many position and momentum statistics via homodyne measurements, and then average these samples to construct the moment matrices \cite[Sec.~3.8.1]{welsch1999ii-homodyne-det} \cite{richter1996determination-o,wunsche1997radon-transform}.

Another method to measure the second moment matrix is given in \cite[Thm.~S53]{mele2024learning-quantu}, and we expect an an analogous method is possible for the fourth moment matrix.

The upshot is that, from \cref{thm:main-thm-const-fock,thm:main-thm-arb-fock,thm:main-thm-gaussian}, it suffices the measure the moment matrices to an error $1/\poly{n, f_{\rm max}, \e^s}$ and this can be done with $\poly{n, f_{\rm max}, \e^s}$ measurements.

For completeness, we perform a full analysis on using heterodyne measurements to learn the moment matrices.
This therefore gives a full, sample and time efficient, end to end algorithm with sample complexity guarantees shown in \cref{cor:nsamples-all1s,cor:nsamples-passive,cor:nsamples-active}.
We emphasize that any measurement technique that learns the moment matrices can then be applied to \cref{thm:main-thm-const-fock,thm:main-thm-arb-fock,thm:main-thm-gaussian} to learn the state. 
Our analysis in this section with heterodyne measurements is just one such choice.

\subsection{Passive Gaussian unitaries}

\begin{proposition}
    \label{prop:heterodyne-measure-passive}
    Let $\ket\psi = \calU_W\ket{\bm f}$ for an unknown unitary $W \in \U(n)$ specifying an arbitrary passive Gaussian unitary and an arbitrary Fock state $\ket{\bm f}$.
    The moment matrix $\sigma^{(t)}$ can be estimated to norm precision $\varepsilon$, $\norm*{\sigma^{(t)\prime} - \sigma^{(t)}} \leq \varepsilon$, with probability at least $1-\delta$ by using $N \sim \bigO{t\log(n/\delta) (2t)! f_{\rm max}^{2t} n^{2t} / \varepsilon^2}$ heterodyne samples,
    where $f_{\rm max} = \max_i f_i$.
\end{proposition}
\begin{proof}
    Denote the heterodyne probability density function as 
    \begin{equation}
        p(\bm\alpha) = \frac{1}{\pi^n}\abs*{\bra{\bm\alpha}\ket\psi}^2, \qquad \bm\alpha \in \bbC^n.
    \end{equation}
    We first note that
    \begin{salign}[eq:heterodyne-expectation]
        \EX{\bm\alpha \sim p}{\alpha_{i_1}\dots \alpha_{i_t}\bar\alpha_{j_1}\dots \bar\alpha_{j_t}}
        &= \int \Dd{n}{\bm\alpha}~p(\bm\alpha)
        \alpha_{i_1}\dots \alpha_{i_t}\bar\alpha_{j_1}\dots \bar\alpha_{j_t} \\
        (\rho = \ket\psi\bra\psi)
        &= \frac{1}{\pi^n}\int \Dd{n}{\bm\alpha}~ \bra{\bm\alpha}\rho \ket{\bm\alpha}
        \alpha_{i_1}\dots \alpha_{i_t}\bar\alpha_{j_1}\dots \bar\alpha_{j_t} \\
        &= \frac{1}{\pi^n}\int \Dd{n}{\bm\alpha}~ \bra{\bm\alpha}a_{j_1}^\dag \dots a_{j_t}^\dag \rho a_{i_1}\dots a_{i_t} \ket{\bm\alpha} \\
        &= \Tr\bargs{a_{j_1}^\dag \dots a_{j_t}^\dag \rho a_{i_1}\dots a_{i_t}} \\
        &= \Tr\bargs{\rho a_{i_1}\dots a_{i_t} a_{j_1}^\dag \dots a_{j_t}^\dag} \\
        &= \bra\psi a_{i_1}\dots a_{i_t} a_{j_1}^\dag \dots a_{j_t}^\dag \ket\psi \\
        &= \sigma^{(t)}_{i_1,\dots,i_t;j_1,\dots,j_t} ~.
    \end{salign}
    Furthermore,
    \begin{salign}
        \variance_{\bm\alpha\sim p}(\alpha_{i_1}\dots \alpha_{i_t}\bar\alpha_{j_1}\dots \bar\alpha_{j_t})
        &\leq \EX{\bm\alpha\sim p}{\abs{\alpha_{i_1}\dots \alpha_{i_t}\bar\alpha_{j_1}\dots \bar\alpha_{j_t}}^2} \\
        &= \bra{\psi}a_{i_1}\dots a_{i_t} a_{j_1} \dots a_{j_t} a_{i_1}^\dag\dots a_{i_t}^\dag a_{j_1}^\dag \dots a_{j_t}^\dag \ket\psi \\
        &= \sigma^{(2t)}_{i_1,\dots,i_t,j_1,\dots,j_t; i_1,\dots,i_t,j_1,\dots,j_t} \\
        &\leq \norm*{\sigma^{(2t)}}_{\rm max} \\
        &\leq \norm*{\sigma^{(2t)}} \\
        &= \norm*{W^{\otimes (2t)}\sigma^{(2t)}_0 W^{\dag \otimes(2t)}} \label{eq:passive1}\\
        &= \norm*{\sigma^{(2t)}_0} \label{eq:passive2} \\
        (\text{Gershgorin circle theorem})
        &\leq 
        \max_{k_1,\dots,k_{2t}}\sum_{k_1',\dots,k_{2t}'} \abs{(\sigma^{(2t)}_0)_{k_1,\dots,k_{2t}; k_1',\dots,k_{2t}'}} \\
        (\ket{\bm f}\text{ is a Fock state})
        &\leq 
        \max_{k_1,\dots,k_{2t}}\sum_{\pi\in S_{2t}} \abs{(\sigma^{(2t)}_0)_{k_1,\dots,k_{2t}; k_{\sigma(1)},\dots,k_{\sigma(2t)}}} \\
        &\leq (2t)! f_{\rm max}^{2t} ~ .
    \end{salign}

    Now that we know the mean and variance, we can construct and analyze an estimator in the setting where
    we sample $N$ times from $p$, resulting in $N$ vectors $S = \set*{\bm\alpha^{(1)}, \dots, \bm\alpha^{(N)}}$.
    Our goal is to build an estimate $\sigma^{(t)\prime}$ from $S$.
    
    We will use a median of means estimator \cite{huang2020predicting-many}, and denote it by $\sigma^{(t)\prime}_{i_1,\dots,i_t; j_1,\dots,j_t}$.
    Given the mean and variance calculations from above, 
    it is immediate \cite{huang2020predicting-many} that $N \sim \bigO{K (2t)! f_{\rm max}^{2t} / \varepsilon^2}$ samples suffice in order to satisfy
    \begin{equation}
        \label{eq:mom-prob-bound}
        \Pr{}{\abs{\sigma^{(t)\prime}_{i_1,\dots,i_t; j_1,\dots,j_t} - \sigma^{(t)}_{i_1,\dots,i_t; j_1,\dots,j_t}} \geq \varepsilon} \leq 2\e^{-K/2}.
    \end{equation}
    It follows that with
    $N \sim \bigO{K (2t)! f_{\rm max}^{2t} n^{2t} / \varepsilon^2}$
    heterodyne samples,
    \begin{salign}
        \Pr{}{\norm*{\sigma^{(t)\prime} - \sigma^{(t)}} \geq \varepsilon} \\
        (\text{use $\norm{A} \leq m\norm{A}_{\rm max}$ for $m\times m$ matrices})
        &\leq 
        \Pr{}{\norm*{\sigma^{(t)\prime} - \sigma^{(t)}}_{\rm max} \geq \frac{\varepsilon}{n^t}} \\
        (\text{union bound})
        &\leq \sum_{i_1,\dots,i_t, j_1,\dots,j_t}
        \Pr{}{\abs{\sigma^{(t)\prime}_{i_1,\dots,i_t; j_1,\dots,j_t} - \sigma^{(t)}_{i_1,\dots,i_t; j_1,\dots,j_t}} \geq \frac{\varepsilon}{n^t}} \\
        (\text{\cref{eq:mom-prob-bound}})
        &\leq 
        2n^{2t}\e^{-K/2}.
    \end{salign} 
    In order for this failure probability to be $\leq \delta$,
    we therefore must set $K = \bigO{t\log(n/\delta)}$.
    Plugging this into $
    N \sim \bigO{K (2t)! f_{\rm max}^{2t} n^{2t} / \varepsilon^2}$
    completes the proof.
\end{proof}

Combining the above measurement bounds with our algorithms yields the full end-to-end sample complexity.

\begin{corollary}
    \label{cor:nsamples-all1s}
    Let $\ket\psi = \calU_W \ket{1^n}$ for an unknown unitary $W \in \U(n)$ specifying an arbitrary passive Gaussian unitary.
    $N \sim \bigO{n^8 \log(n/\delta) / \varepsilon^2}$ heterodyne samples from $\ket\psi$ suffice to learn, via the efficient \cref{alg:main-alg-const-fock}, a $V \in \U(n)$ such that $\abs*{\bra{1^n}\calU_W^\dag \calU_V\ket{1^n}} \geq 1 - \varepsilon$ with probability at least $1-\delta$.
\end{corollary}
\begin{proof}
    Use $N$ heterodyne samples via \cref{prop:heterodyne-measure-passive} to build an estimate $\sigma^{(2)\prime}$ of 
    $\sigma^{(2)}$ satisfying $\norm*{\sigma^{(2)\prime} - \sigma^{(2)}} \leq \bigO{\varepsilon / n^2}$ with probability $1-\delta$.
    Then run \cref{alg:main-alg-const-fock}, which, by \cref{thm:main-thm-const-fock}, 
    finds $V$ such that 
    $\abs*{\bra{1^n}\calU_W^\dag \calU_V\ket{1^n}} \geq 1-\bigO{\varepsilon}$.
\end{proof}

Note that in the numerics in \cref{fig:numerics} where we analyze exactly the setting of \cref{cor:nsamples-all1s}, we do not simulate the median of means estimator but rather just the empirical mean estimator. Nonetheless, numerically we find a scaling of $N\sim n^6$, which is better than the $\sim n^8\log n$ scaling proved in \cref{cor:nsamples-all1s}.

Next we consider the case where the initial state can be an arbitrary Fock state.

\begin{corollary}
    \label{cor:nsamples-passive}
    Let $\ket\psi = \calU_W \ket{\bm f}$ for an unknown unitary $W \in \U(n)$ specifying an arbitrary passive Gaussian unitary and an arbitrary unknown Fock state specified by $\bm f = (f_1,\dots,f_n)$.
    Let $f_{\rm max} = \max_i f_i$.
    $N \sim \bigO{n^{10} f_{\rm max}^{10} \log(n/\delta) / \varepsilon^2}$
    heterodyne samples from $\ket\psi$ suffice to learn, via the efficient \cref{alg:main-alg-arb-fock}, a $V \in \U(n)$ and $\bm g$ such that $\abs*{\bra{\bm f}\calU_W^\dag \calU_V\ket{\bm g}} \geq 1 - \varepsilon$ with probability at least $1-\delta$.
\end{corollary}
\begin{proof}
    Via \cref{prop:heterodyne-measure-passive},
    using $\bigO{n^4 f_{\rm max}^{4} \log(n/\delta) / \tilde\varepsilon^2}$ suffices to estimate the moment matrices $\sigma^{(1)},\sigma^{(2)}$ to norm precision $\tilde\varepsilon$.
    From \cref{thm:main-thm-arb-fock},
    in order to achieve the desired infidelity of $\varepsilon$, it suffices that we satisfy $\tilde\varepsilon \sim \frac{\varepsilon}{n^3 f_{\rm max}^3}$.
    Plugging this in completes the proof.
\end{proof}

We note that 
the $n$ dependence is actually worse in \cref{cor:nsamples-passive} than in \cref{cor:nsamples-active} below.
This is an artifact of the $n$ dependence in \cref{eq:block-distance-arb-fock} which is not tight.
Of course, \cref{cor:nsamples-active} can be applied to \cref{cor:nsamples-passive}, but we still include \cref{cor:nsamples-passive} as it slightly more direct.

\subsection{Arbitrary Gaussian unitaries}

\begin{proposition}
    \label{prop:heterodyne-measure-gaussian}
    Let $\ket\psi = \calU_S \ket{\bm f}$ for an unknown symplectic matrix $S \in \Sp(2n, \bbR)$ specifying an arbitrary Gaussian unitary and an arbitrary Fock state $\ket{\bm f}$.
    The moment matrix $\Lambda^{(t)}$ can be estimated to norm precision $\varepsilon$, $\norm*{\Lambda^{(t)\prime} - \Lambda^{(t)}} \leq \varepsilon$, with probability at least $1-\delta$ by using 
    $N \sim \bigO{t\log(n/\delta) (2t)! f_{\rm max}^{2t} \e^{4ts} n^{2t} / \varepsilon^2}$
    heterodyne samples,
    where $f_{\rm max} = \max_i f_i$ and $\e^{2s} = \norm*{S^TS}$.
\end{proposition}
\begin{proof}
    Firstly, rather than considering $\Lambda^{(t)}_{i_1,\dots,i_t; j_1,\dots,j_t} = \bra\psi r_{i_1}\dots r_{i_t}r_{j_1}\dots r_{j_t}\ket\psi$ for $r_i = x_i$ and $r_{n+i} = p_i$, we will consider the unitarily rotated basis so that
    $\tilde\Lambda^{(t)}_{i_1,\dots,i_t; j_1,\dots,j_t} = \bra\psi q_{i_1}\dots q_{i_t}q_{j_1}\dots q_{j_t}\ket\psi$, where $q_i = a_i$ and $q_{n+i} = a_i^\dag$.
    The unitary transformation comes from \cref{eq:a-adag-x-p}.
    Because this unitary transformation is so simple, the result error analysis is unchanged.
    The anti-normal ordered elements of $\tilde\Lambda^{(t)}$ can be estimated directly by averaging heterodyne samples (as in \cref{prop:heterodyne-measure-passive}), and the remaining moments are computed from the anti-normal ordered moments along with computation (see our code \cite{iosuegithub} for an explicit algorithm that does this).

    Now the proof is almost exactly the same as that of \cref{prop:heterodyne-measure-passive}.
    The difference is in going from \cref{eq:passive1} to \cref{eq:passive2}.
    In place of $\norm*{W^{\otimes (2t)}\sigma_0^{(2t)} W^{\dag \otimes (2t)}} = \norm*{\sigma^{(2t)}_0}$,
    we have
    \begin{equation}
        \norm*{S^{\otimes (2t)} \tilde\Lambda^{(2t)}_0 S^{T \otimes (2t)}} \leq \e^{4ts} \norm*{\tilde\Lambda^{(2t)}_0},
    \end{equation}
    where $\e^{s}$ is the maximum singular value of $S$ (that is, $\e^{2s} = \norm*{S^T S}$).

    Tracking this change through the proof of \cref{prop:heterodyne-measure-passive} completes this proof.
\end{proof}

\begin{corollary}[Corollary 3 in the main text]
    \label{cor:nsamples-active}
    Let $\ket\psi = \calU_S \ket{\bm f}$ for an unknown symplectic matrix $S \in \Sp(2n, \bbR)$ specifying an arbitrary Gaussian unitary and an arbitrary Fock state $\ket{\bm f}$.
    Let $f_{\rm max} = \max_i f_i$.
    $N \sim \bigO{n^8 f_{\rm max}^{10} \e^{20s} 
    \log(n/\delta) / \varepsilon^2}$
    heterodyne samples from $\ket\psi$ suffice to learn, via the efficient \cref{alg:main-alg-gaussian}, a $Q \in \Sp(2n, \bbR)$ and $\bm g$ such that $\abs*{\bra{\bm f}\calU_S^\dag \calU_Q\ket{\bm g}} \geq 1 - \varepsilon$ with probability at least $1-\delta$.
\end{corollary}
\begin{proof}
    Via \cref{prop:heterodyne-measure-gaussian},
    using $\bigO{n^4 f_{\rm max}^{4} \e^{8s} \log(n/\delta) / \tilde\varepsilon^2}$ suffices to estimate the moment matrices $\Lambda^{(1)},\Lambda^{(2)}$ to norm precision $\tilde\varepsilon$.
    From \cref{thm:main-thm-gaussian},
    in order to achieve the desired infidelity of $\varepsilon$, it suffices that we satisfy 
    $\tilde\varepsilon n^2\e^{6s}f_{\rm max}^{3}\sim \varepsilon$.
    Plugging this in completes the proof.
\end{proof}

We believe these bounds can be made slightly tighter, and that the $\e^s$ dependence can be much better with an adaptive sampling approach \cite{bittel2025energy-independ}.

\cref{cor:nsamples-active} is our full end-to-end result for learning states of the form $\calU_S\ket{\bm f}$.
We compare our result to other learning algorithms in \cref{tab:comparison}.
No other algorithms exist to efficiently learn such states.

\begin{table*}
\centering
    \begin{tabular}{|p{7.5cm}|c|c|}
        \hline
        \textbf{} & \textbf{Sample complexity} & \textbf{Time complexity} \\
        \hline
        \raggedright Learning $t$-doped Gaussian states \cite{mele2024learning-quantu} & {$\exp(n)$}  & {$\exp(n)$}  \\
        \raggedright  Learning pure ``Gaussian-entanglable'' states via classical shadows \cite{zhao2024complexity-of-q} & $\poly{n}$ & $\exp(n)$\\
        \raggedright \textit{(This work)} Learning BosonSampling states & {$ \poly{n}$} & {$\poly{n}$ } \\
        \hline
    \end{tabular}
    \caption{Summary of algorithms applied to states of the form $\calU_S \vert \bm f\rangle$. We note that our algorithm specifically applies to these states, while the algorithms in \cite{mele2024learning-quantu,zhao2024complexity-of-q} apply to other settings.
    We further note that there are many other learning algorithms for bosonic Gaussian states, but they are not able to learn states of the form $\calU_S\vert\bm f\rangle$ due to the non-Gaussianity.}
    \label{tab:comparison}
\end{table*}

\section{Numerical simulations}
\label{sec:numerics}

\begin{figure}
    \begin{minipage}{.49\textwidth}
    \includegraphics[width=\textwidth]{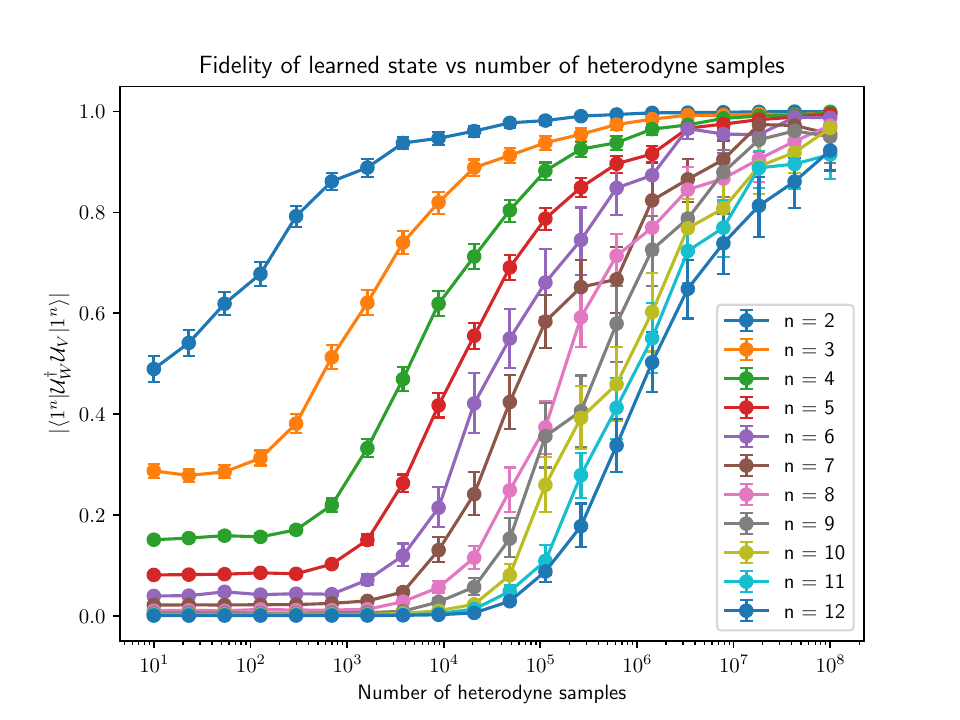}
    \end{minipage}%
    \hfill%
    \begin{minipage}{.49\textwidth}
    \includegraphics[width=\textwidth]{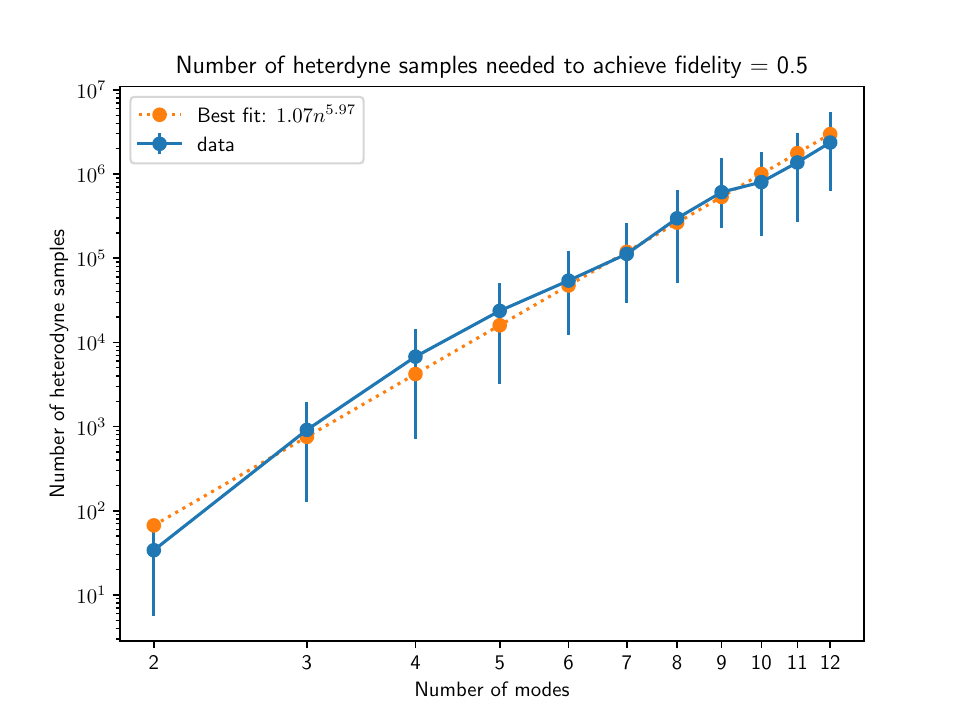}
    \end{minipage}
    \caption{
    Numerical simulations (see our code \cite{iosuegithub}) of the performance of heterodyne sampling with \cref{alg:main-alg-const-fock} for learning states 
    $\mathcal U_W \vert 1^n\rangle$ for random $W$.
    (Left) The fidelity $\vert\langle1^n\vert\calU_W^\dag \calU_V \vert1^n\rangle\vert$ of the learned state with the true state after running \cref{alg:main-alg-const-fock} with correlation matrix $\sigma^{(2)}$ empirically estimated from $N$ heterodyne samples.
    (Right) The number of heterodyne samples needed for \cref{alg:main-alg-const-fock} to learn a $V$ that achieves a fidelity $\vert\langle1^n\vert\calU_W^\dag \calU_V \vert1^n\rangle\vert \geq 1/2$.
    For both plots, the error bars are computed using the bootstrap method, where we 
    resample the data with replacement, recompute the mean, take the 2.5\% and 97.5\% quantiles.
    }
    \label{fig:numerics}
\end{figure}

In this section, we numerically simulate heterodyne sampling from a state $\calU_W\ket{1^n}$ and running \cref{alg:main-alg-const-fock} to learn a $V$, and computing the overlap $\abs*{\bra{1^n}\calU_W^\dag\calU_V\ket{1^n}}$.
We describe how the simulations are performed in more detail below, but the summary is as follows.
We generate heterodyne samples from $N$ copies of the state $\calU_W \ket{1^n}$ and use them to approximate the correlation matrix $\sigma^{(2)}$.
We then run our \cref{alg:main-alg-const-fock} via \texttt{\sc findV}$(\sigma^{(2)}, 1)$ to find the unitary $V$. Finally, we compute the overlap $\abs*{\bra{1^n}\calU_W^\dag\calU_V\ket{1^n}}$.
We do this for many different random $W$ and average the results.
\cref{fig:numerics} (left) shows the overlap as a function of $N$.
Next, we define $N^\ast$ as the smallest value of $N$ needed to achieve an overlap of at least $1/2$.
\cref{fig:numerics} (right) shows $N^\ast$ vs the number of modes $n$.
As can be seen from \cref{fig:numerics} (right), when drawing random $W$, the typical number of heterodyne samples needed to achieve an overlap of $1/2$ scales as $\sim n^6$.
This is a substantial improvement over the worst case rigorous bound of $\sim n^{8}\log n$ that we derived in \cref{cor:nsamples-all1s}.

\textbf{Simulation details.}~
All of our code and simulations is available on GitHub \cite{iosuegithub}.
\cref{alg:main-alg-const-fock} is simply implemented as described with standard linear algebra tools.
The overlaps are computed via permanents with The Walrus \cite{thewalrus}.

The main nontrivial part of the simulation is thus, given a unitary $W\in \U(n)$, to generate $N$ heterodyne samples from the state $\calU_W \ket{1^n}$.
This means we must generate samples from the probability distribution $p(\bm \alpha) = p(\alpha_1,\dots,\alpha_n) = \frac{1}{\pi^n} \abs{\bra{\bm\alpha} \calU_W \ket{1^n}}^2$, where $\bm \alpha \in \bbC^n$.
We have that 
\begin{equation}
    \label{eq:heterodyne-distribution}
    p(\bm \alpha)
    =\frac{1}{\pi^n} \abs{\bra{\bm\alpha} \calU_W \ket{1^n}}^2 
    =\frac{1}{\pi^n} \abs{\bra{W^\dag \bm\alpha}\ket{1^n}}^2 .
\end{equation}
We therefore only need to generate samples $\beta_i \sim \abs{\bra{\beta_i}\ket1}^2 \propto \e^{-\abs*{\beta_i}^2} \beta_i^2$,
form them into a vector $\bm \beta = (\beta_1,\dots,\beta_n)$, and return the sample $\bm\alpha = W \bm \beta$.
The distribution $\e^{-\abs*{\beta_i}^2} \beta_i^2$ is simply a Gamma distribution for its magnitude, and a uniform distribution for its phase.
Numerically, we use NumPy to sample from the Gamma distribution.

Once we have the $N$ heterodyne samples $\bm \alpha^{(1)}, \dots, \bm \alpha^{(N)}$, we estimate the correlation matrix entry $\sigma^{(2)}_{i,j;k,\ell} = \angles*{a_i a_j a_k^\dag a_\ell^\dag}$ as 
$\frac{1}{N}\sum_{m=1}^N \alpha^{(m)}_i \alpha^{(m)}_j \bar\alpha^{(m)}_k \bar\alpha^{(m)}_\ell$.
This comes from the calculation performed in \cref{eq:heterodyne-expectation}.
So, with finitely many samples, we replace the integral with an empirical average.
Note that our measurement bounds in \cref{cor:nsamples-all1s}, we show that the number of required measurements $N^\ast$ scales at most as $n^8 \log n$ when using a median of means estimator.
For the numerical simulations for simplicity, we do not use a median of means estimator but rather just a standard empirical average estimator.
Nonetheless, we we find $N^\ast$ to numerically scale as only $n^6$ in \cref{fig:numerics}.

Numerically, most of this simulation is efficient.
For a given $W$, numerically generating heterodyne samples from $\calU_W\ket{1^n}$, estimating $\sigma^{(2)}$ via an empirical average, and running our algorithm \texttt{\sc findV} are all efficient.
However, computing the overlap $\abs*{\bra{1^n}\calU_W^\dag \calU_V \ket{1^n}} = \abs*{\perm(W^\dag V)}$
is not numerically efficient.
Therefore, we only run the simulations up to $n=12$ modes.

\ignore{
\section{\texorpdfstring{$G_t$}{Gt} states are defined by their first \texorpdfstring{$t$}{t} moments}
\label{sec:Gt-states}

In the main text, we noted that
a \Gt state is fully specified by its first $t$ moments.
We now provide more details.

We begin with a mixed \Gt state $\rho$.
Recall that a mixed \Gt state is a thermal state of a degree $\leq t$ Hamiltonian.
Let $\hat M_1,\hat M_2,\dots$ be all the moment operators up to degree $t$, and let $M_i = \Tr[\rho \hat M_i]$.
Suppose Alice gives Bob the moments $M_i$ for all $i$, and Alice promises Bob that those moments came from a \Gt state.
Note that Bob has no access to or knowledge of $\rho$ besides what Alice told him.
Because of the promise, Bob knows that $\rho$ is a Gibbs state of a degree $\leq t$ Hamiltonian.
Using Ref.~\cite{jaynes1957information-the-II}, it follows that $\rho$ is the maximal entropy state subject to constraints on its first $t$ moments. Thus, Bob in principle has enough information to completely reconstruct the state, as he can then perform the following maximization,
\begin{equation}
    \begin{split}
        \rho = \max_\sigma &\parentheses{-\Tr\bargs{\sigma\log\sigma}} \\
        &\text{s.t.~} \forall i\colon~ \Tr[\sigma \hat M_i] = M_i
        .
    \end{split}
\end{equation}

Next, we consider the pure state case.
Recall a pure \Gt state $\psi$ is the unique ground state of a non-degenerate Hamiltonian $H$.
We now want to show that such a state is uniquely specified by its first $t$ moments.
$H$ can be written as $\sum_i c_i \hat M_i$.
Then we have
\begin{salign}
    \bra\psi H \ket\psi
    &= \min_{\ket\phi} \sum_i c_i \bra\phi \hat M_i \ket\phi \\
    \begin{split}
    &= \min_{m_1,m_2,\dots \in \bbC}\sum_i c_i m_i \\
    &\qquad \text{s.t. there exists a state $\phi$ satisfying } \bra\phi \hat M_i \ket\phi = m_i ~ \forall i 
    \end{split} .
\end{salign}
Let $M_i$ be the minimizing $m_i$ (eg.~change min to argmin).
Notice that the $\phi$ that satisfies $\bra\phi \hat M_i \ket\phi = M_i$ is uniquely the ground state of $H$, and so $\ket\phi = \ket\psi$.

Now, as before, suppose that Alice gives Bob the numbers $M_i \in \bbC$ and promises that $M_i = \bra\psi \hat M_i \ket\psi$ for a \Gt state $\psi$, but Bob knows nothing else about $\psi$.
With only this information, Bob can in principle reconstruct the state, because from above, he is guaranteed that 
$\ket\psi$ is the unique state that satisfies $M_i = \bra\psi \hat M_i \ket\psi$ for all $i$.

\paragraph*{Discussion.}%
A \Gt state is fully specified by its first $t$ moments in the sense above.
Importantly, we have assumed that the first $t$ moments are known \textit{exactly}. 
If the moments are only known \textit{approximately}, then it is an open and interesting question how accurately one needs to know the first $t$ moments in order to be able to in principle reconstruct the state to a desired fidelity.
Indeed, our learning algorithm in \cref{thm:main-thm-gaussian} is precisely an answer to this question in the case of a restricted class of \G*4 states, namely Fock states acted upon by Gaussian unitaries.
Similarly, Ref.~\cite{mele2024learning-quantu} solves this problem in the case of $t=2$ (i.e.~Gaussian states).
}

\bibliographystyle{apsrev4-2}
\bibliography{references}